\documentclass[1p]{elsarticle}
\pdfoutput=1  

\usepackage{lineno}
\usepackage{amsmath,amsfonts}
\usepackage{graphicx}
\usepackage{fullpage}
\usepackage{algorithm}
\usepackage{algpseudocode}
\usepackage{pifont}
\newcommand{\mydraft}{false}
\newcommand{\mb}[1]{\mathbf{#1}}
\newcommand{\gp}[1]{{\left({#1}\right)}}
\newcommand{\mx}[1]{\begin{pmatrix}#1\end{pmatrix}}
\newcommand{\xx}{\mb{x}}
\newcommand{\XX}{\mb{X}}

\newcommand{\uu}{\mb{u}}
\newcommand{\rr}{\mb{r}}
\newcommand{\vv}{\mb{v}}
\newcommand{\VV}{\mb{V}}
\newcommand{\dt}{\Delta t}
\newcommand{\dx}{\Delta x}

\newcommand{\GG}{\mb{G}}
\renewcommand{\SS}{\mb{S}}
\newcommand{\pp}{\mb{p}}
\renewcommand{\ss}{\mb{s}}

\newcommand{\op}{\mbox{\boldmath{$\omega$}}_p}

\newcommand{\ep}{\mbox{\boldmath{$\epsilon$}}}
\newcommand{\sg}{\mbox{\boldmath{$\sigma$}}}
\newcommand{\ph}{\mbox{\boldmath{$\phi$}}}

\renewcommand{\AA}{\mb{A}}
\renewcommand{\ll}{\mb{l}}
\newcommand{\CC}{\mb{C}}
\newcommand{\PP}{\mb{P}}

\newcommand{\MM}{\mb{M}}

\newcommand{\BB}{\mb{B}}
\newcommand{\bb}{\mb{b}}

\newcommand{\DD}{\mb{D}}
\newcommand{\dv}{\Delta\mb{v}}
\newcommand{\yy}{\mb{y}}

\newcommand{\OO}{\mb{O}}
\newcommand{\FF}{\mb{F}}
\newcommand{\ff}{\mb{f}}
\newcommand{\zz}{\mb{z}}
\newcommand{\ZZ}{\mb{Z}}

\newcommand{\vt}{\tilde{\vv}}
\newcommand{\xt}{\tilde{\xx}}

\newcommand{\KK}{\mb{K}}

\newcommand{\II}{\mb{I}}
\newcommand{\z}{\mb{0}}

\newcommand{\ee}{\mb{e}}

\renewcommand{\gg}{\mb{g}}
\newcommand{\norm}[1]{\left\lVert#1\right\rVert}
\newcommand{\px}[2]{\frac{\partial #1}{\partial #2}}
\modulolinenumbers[5]

\journal{Journal of Computational Physics}

\begin{document}

\begin{frontmatter}

\title{An angular momentum conserving Affine-Particle-In-Cell method}

\author{Chenfanfu Jiang}
\author{Craig Schroeder}
\author{Joseph Teran\fnref{myfootnote}}
\fntext[myfootnote]{jteran@math.ucla.edu}
\address{Department of Mathematics\\University of California Los Angeles}

\begin{abstract}
We present a new technique for transferring momentum and velocity between particles and
grid with Particle-In-Cell (PIC) \cite{harlow:1964:pic} calculations which we call
Affine-Particle-In-Cell (APIC). APIC represents particle velocities as locally affine,
rather than locally constant as in traditional PIC. We show that this representation
allows APIC to conserve linear and angular momentum across transfers while also
dramatically reducing numerical diffusion usually associated with PIC. Notably,
conservation is achieved with lumped mass, as opposed to the more commonly used Fluid
Implicit Particle (FLIP) \cite{brackbill:1986:flip-pic,brackbill:1988:flip-dissipation}
transfers which require a ``full'' mass matrix for exact conservation. Furthermore, unlike
FLIP, APIC retains a filtering property of the original PIC and thus does not accumulate
velocity modes on particles as FLIP does. In particular, we demonstrate that APIC does not
experience velocity instabilities that are characteristic of FLIP in a number of Material
Point Method (MPM) hyperelasticity calculations. Lastly, we demonstrate that when combined
with the midpoint rule for implicit update of grid momentum that linear and angular
momentum are exactly conserved.
\end{abstract}


\end{frontmatter}

\section{Introduction}

PIC methods have been used for decades to simulate many different physical
phenomena. Examples include compressible flow, incompressible flow, plasma physics,
computational solids and many more \cite{grigoryev:2002:pic_book}. PIC utilizes a hybrid
particle/grid representation of material to retain the accuracy of Lagrangian techniques
without sacrificing the robustness of Eulerian techniques. In all cases, the hybrid nature
of the approach requires the transfer of state to and from Lagrangian particles and
Eulerian grid. Unfortunately, this frequent remapping can introduce significant error and
instability. The most apparent error is excessive dissipation incurred from double
interpolation. The FLIP approach of Brackbill et al.
\cite{brackbill:1986:flip-pic,brackbill:1988:flip-dissipation} was developed to reduce the
dissipation by transferring \emph{changes} in grid quantities to particles, rather than
directly interpolating as in PIC. This also greatly improved the angular momentum
conservation properties of the particle/grid transfers
\cite{burgess:1992:flip-pic,brackbill:1987:compressible-angular}. However, as pointed out
in \cite{love:2006:stable-mpm} exact conservation with FLIP is only possible with the use
of the ``full'' mass matrix. FLIP cannot guarantee exact conservation when used with the
more efficient ``lumped'' mass matrix. Unfortunately, since the full mass matrix can be
singular for certain particle configurations, it is necessary in practice to interpolate
between a mass-lumped and full mass matrix to avoid issues caused by a poorly conditioned
mass matrix \cite{love:2006:stable-mpm}. However, even with mass lumping, FLIP greatly
reduces the angular momentum losses from transfers in the original PIC.

\begin{figure}
\begin{center}
\includegraphics[draft=\mydraft,width=.333\textwidth]{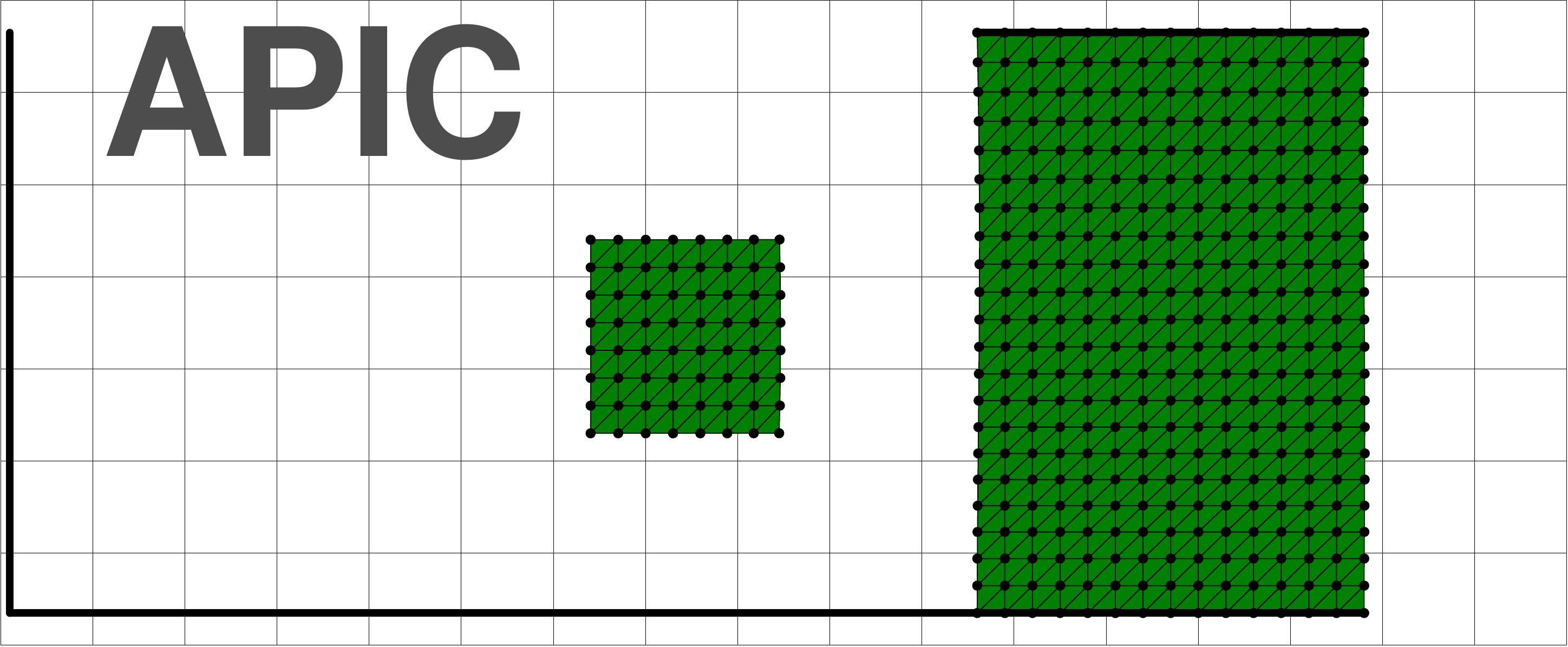}%
\includegraphics[draft=\mydraft,width=.333\textwidth]{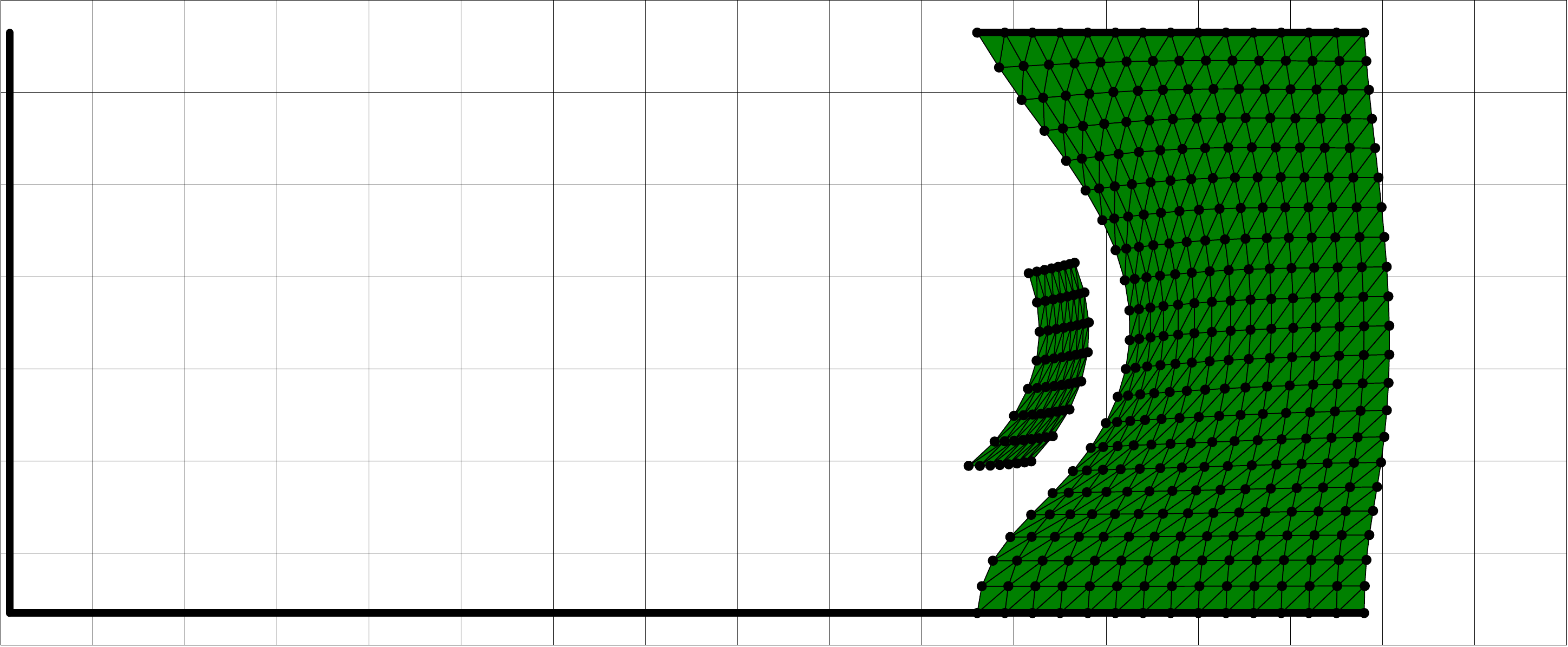}%
\includegraphics[draft=\mydraft,width=.333\textwidth]{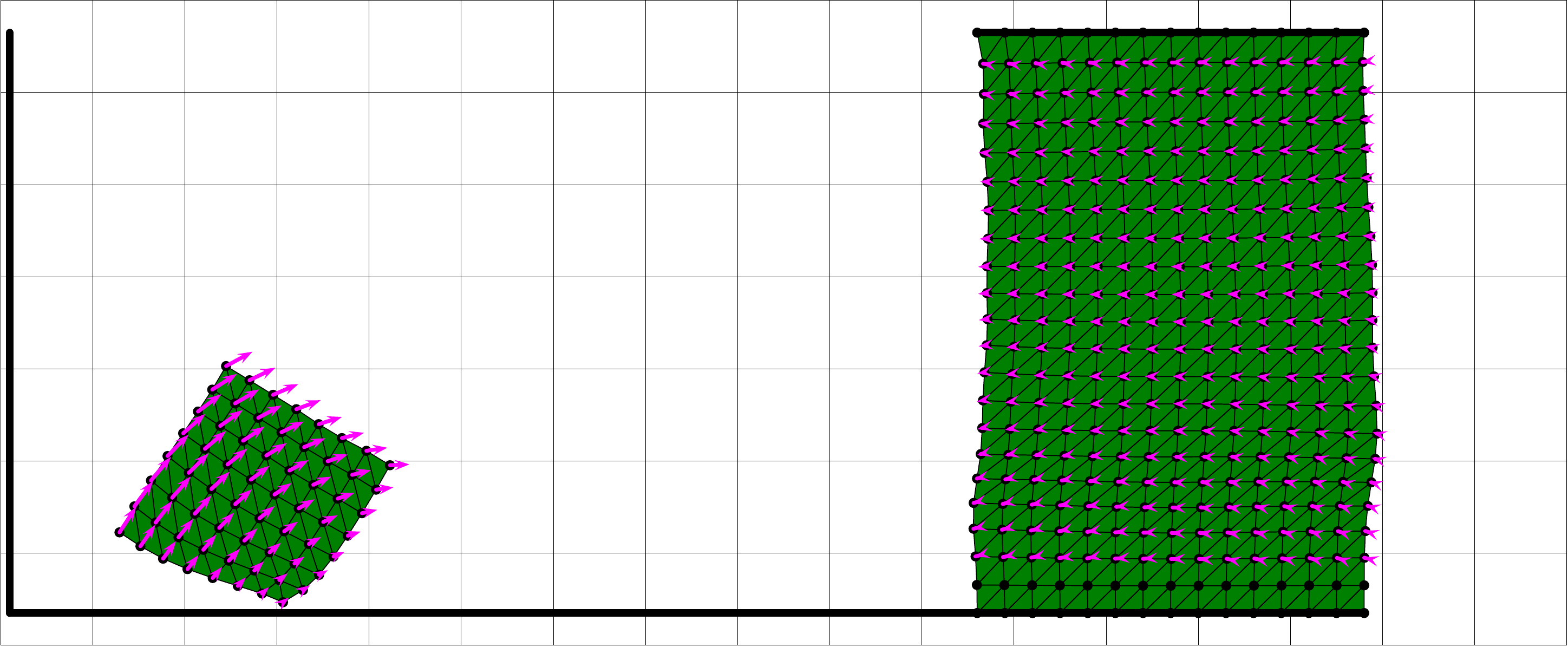}\\*[-1.4em]
\noindent\rule[0cm]{\textwidth}{3pt}
\includegraphics[draft=\mydraft,width=.333\textwidth]{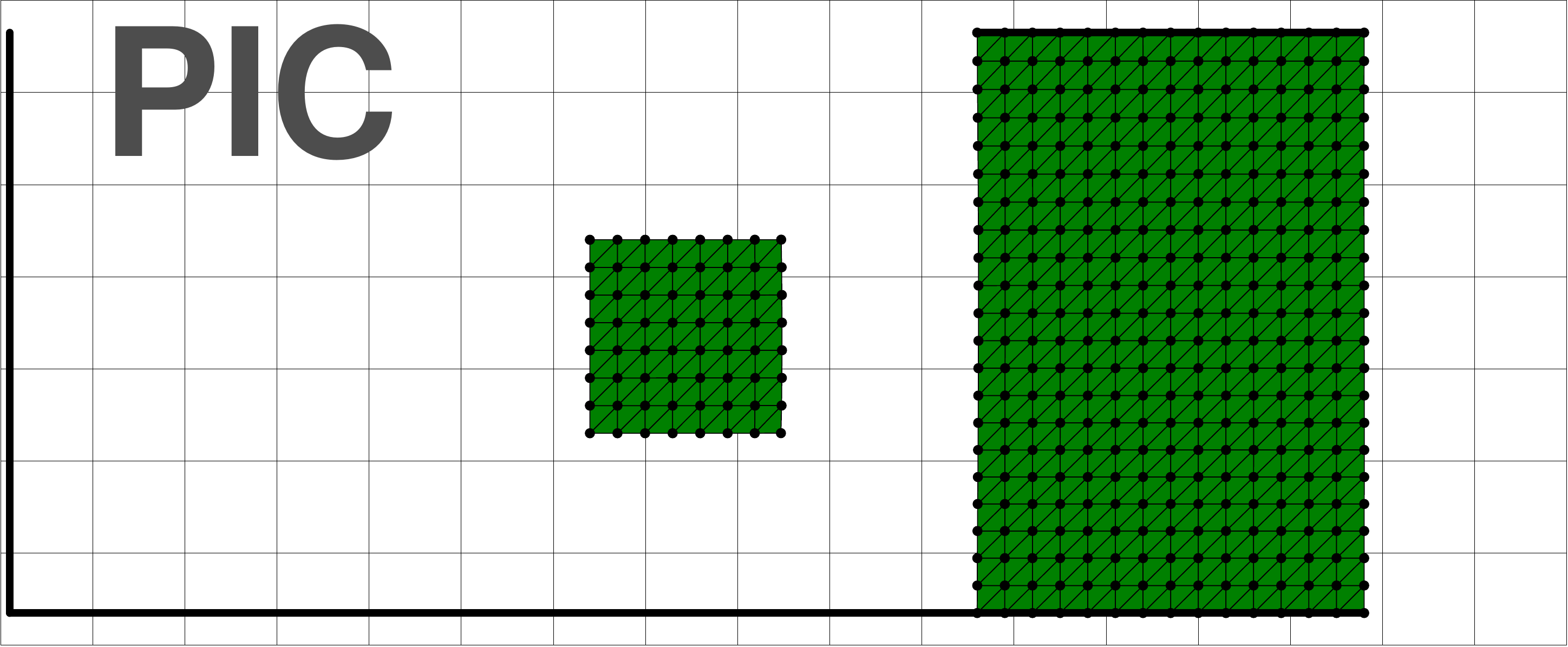}%
\includegraphics[draft=\mydraft,width=.333\textwidth]{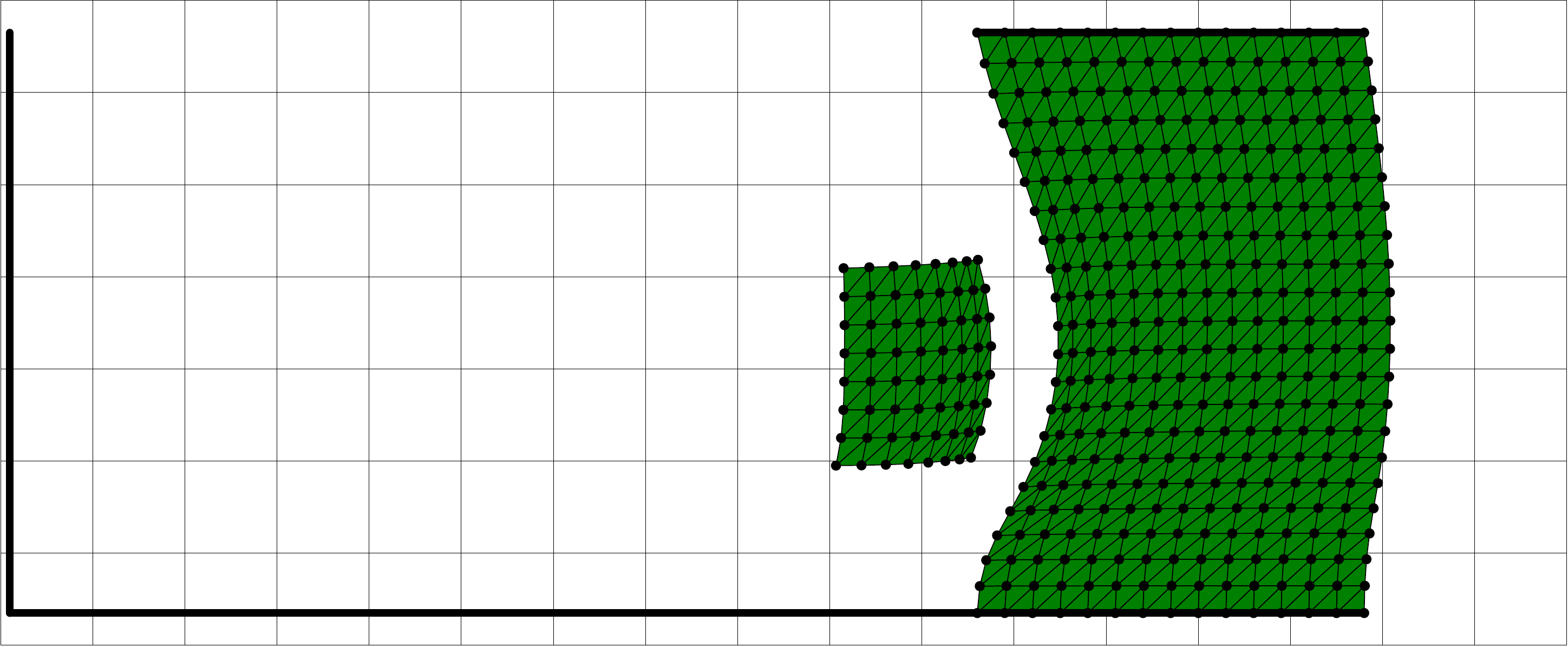}%
\includegraphics[draft=\mydraft,width=.333\textwidth]{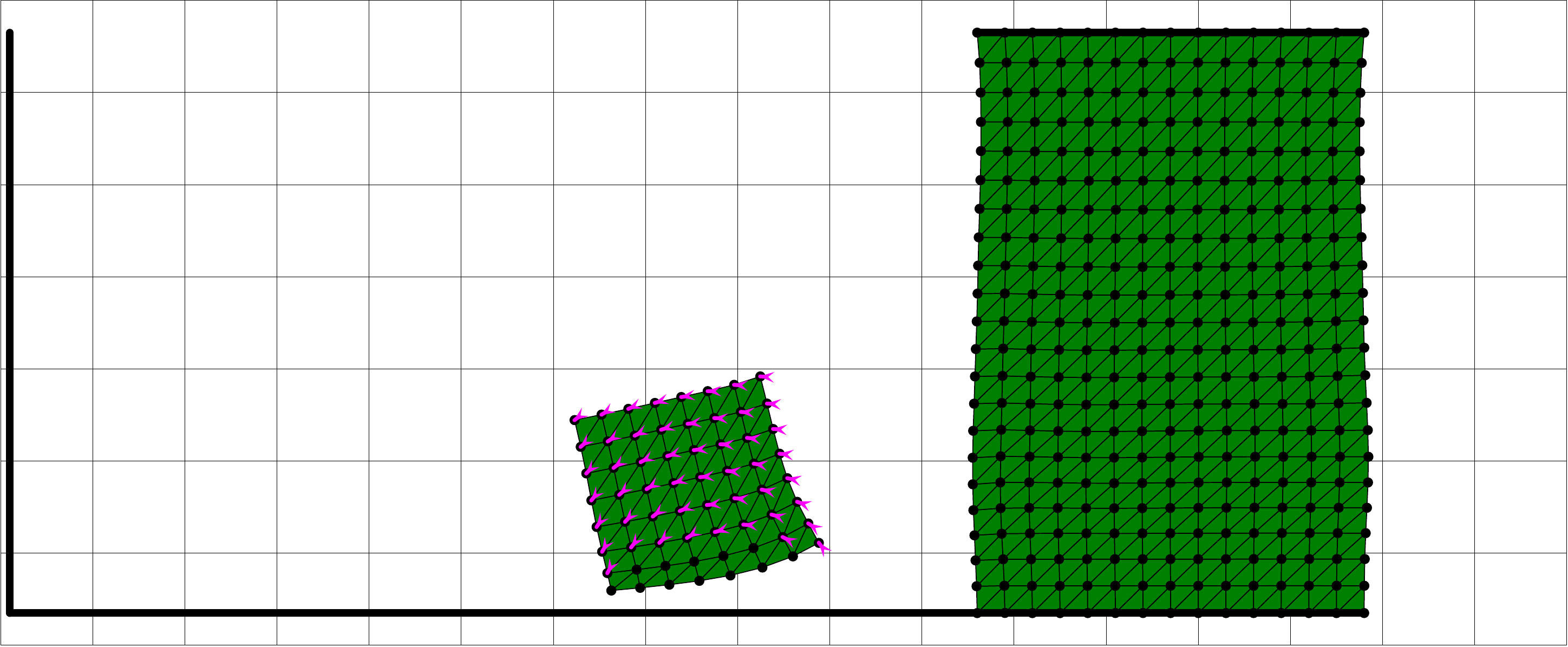}\\*[-1.4em]
\noindent\rule[0cm]{\textwidth}{3pt}
\includegraphics[draft=\mydraft,width=.333\textwidth]{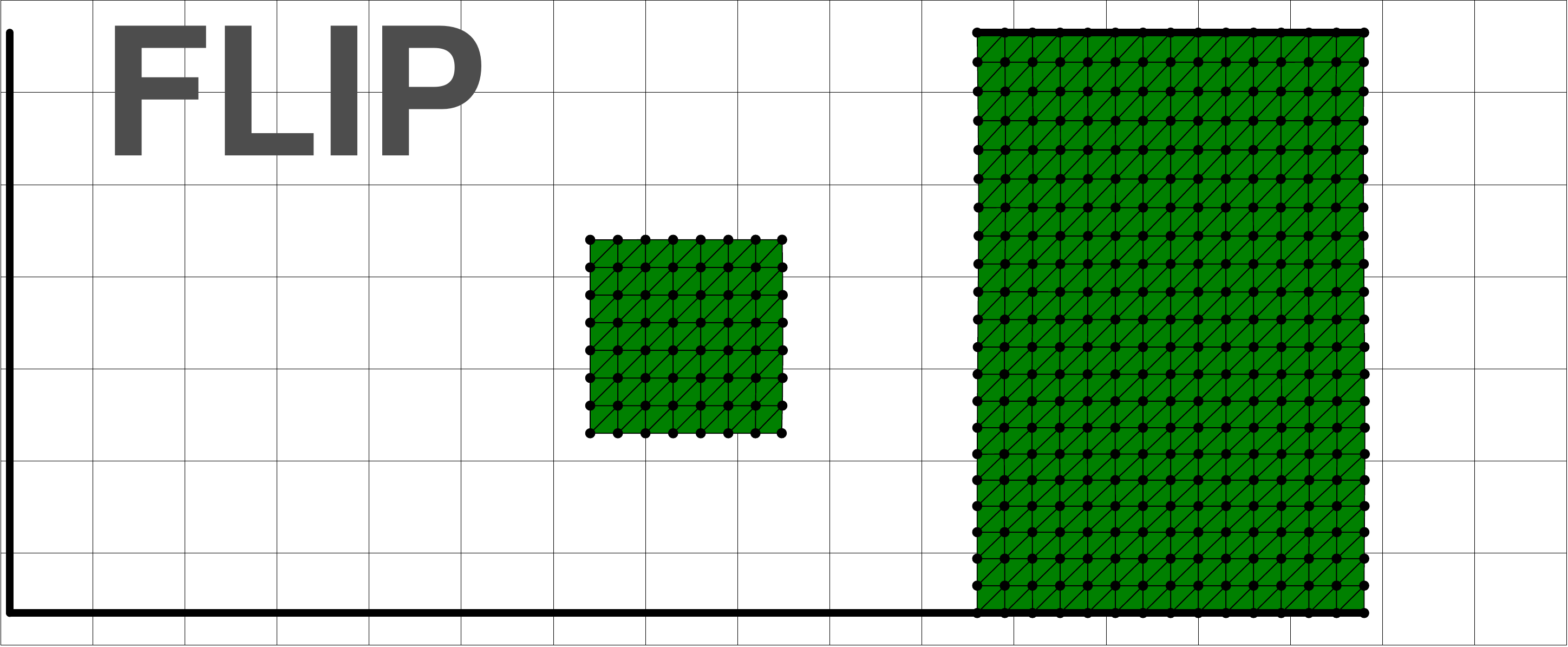}%
\includegraphics[draft=\mydraft,width=.333\textwidth]{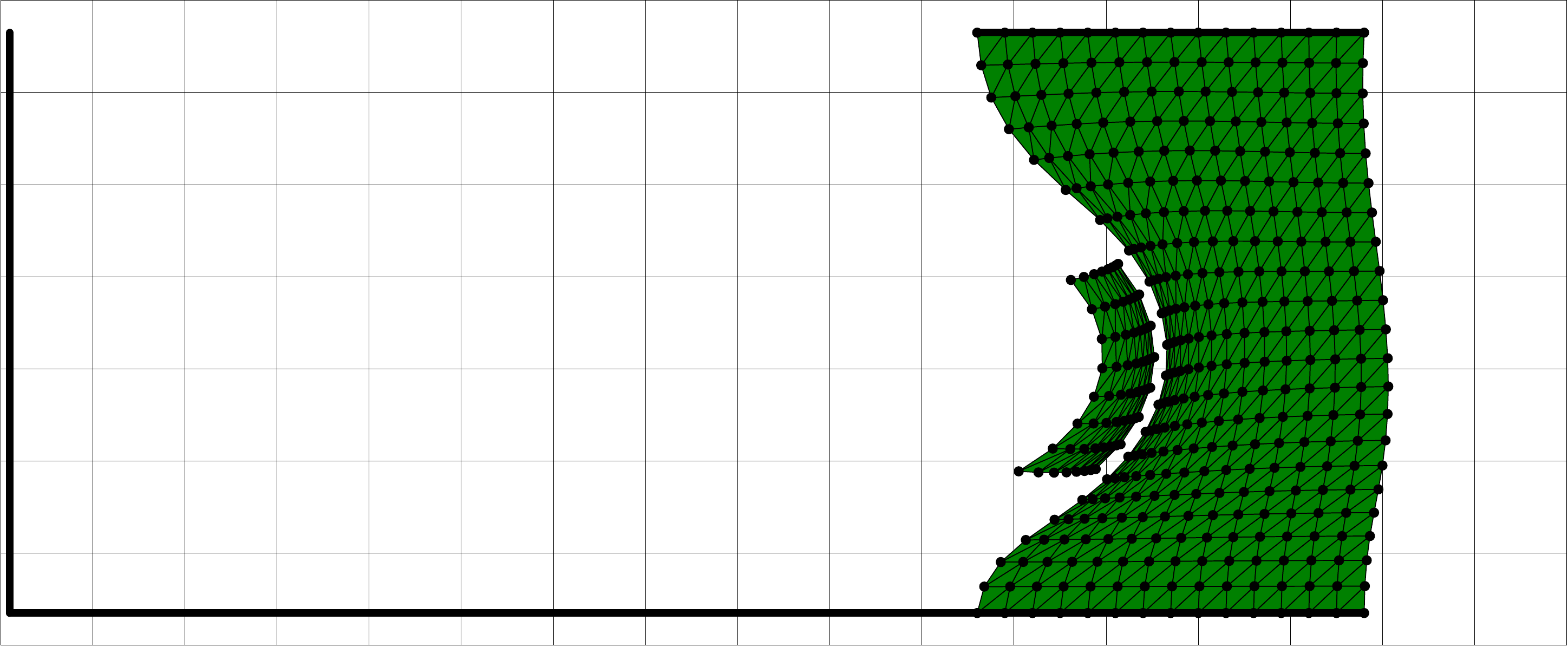}%
\includegraphics[draft=\mydraft,width=.333\textwidth]{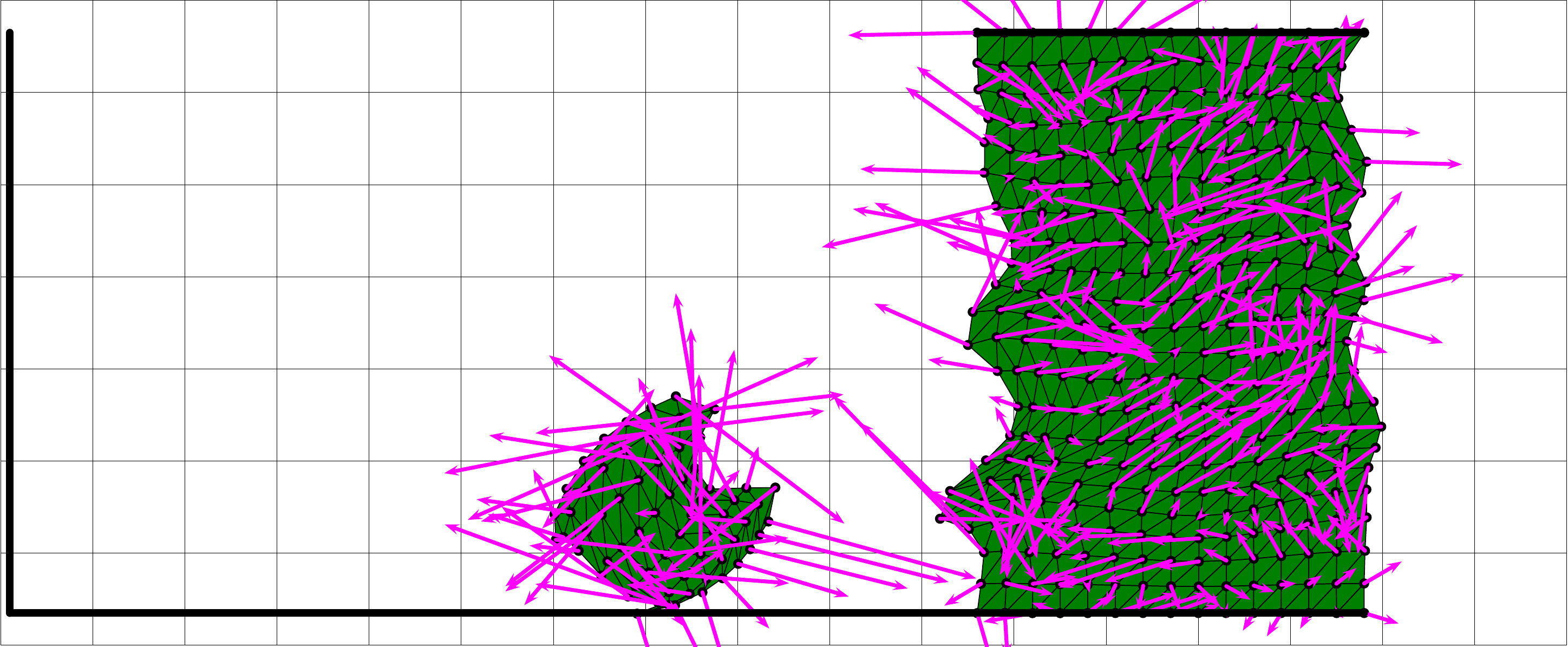}
\end{center}
\caption{Ringing test.  Particle velocities are drawn to illustrate the noisy modes
  persistent with FLIP transfers. PIC and APIC transfers do not suffer from this due to
  the filtering property. However, APIC is not excessively damped like
  PIC. \label{fig:ringing}}
\end{figure}

While all PIC approaches suffer to some degree from finite grid
\cite{langdon:1970:finite_grid,okuda:1972:PIC_inst} (or ringing
\cite{brackbill:1988:ringing-pic,gritton:2014:ringing}) instabilities, FLIP appears to
exacerbate null modes in the transfer operator from particle to grid. This is particularly
true with MPM \cite{sulsky:1994:history-materials,sulsky:1995:pic} PIC techniques for
simulating history dependent materials. The problems arise from the mismatch in particle
and grid degrees of freedom. Typically there are many more particles than grid nodes and
thus information is lost in the particle to grid transfer. While the original PIC
transfers can be seen as a filter of particle degrees of freedom by modes resolvable on
the grid, FLIP does not have this property. FLIP transfers can be shown to cause
unpredictable behavior since certain particle velocity modes persist, invisible to the
dynamics on the grid only to reappear after particle movement. Notably, the particle
velocities are not used to move the particle positions. Particle positions are directly
interpolated from the grid, which is equivalent to using an interpolated, PIC velocity for
position updates. This is idea has also been used for example in
\cite{monaghan:1989:pen}. However, while this reduces the effect of the velocity modes
greatly, it does not completely remove the problem. We illustrate this in
Figure~\ref{fig:ringing}. Despite these issues, FLIP transfers are still most commonly
used, particularly for MPM.

The typical PIC transfer of particle velocities $\vv_p$ to grid velocities $\vv_i$ is done
by first transferring mass and momentum from particle to grid and then dividing out mass
to get velocity as
\begin{equation}
m_i=\sum_pm_pN(\xx_p-\xx_i), \ \ (m\vv)_i=\sum_pm_p\vv_p N(\xx_p-\xx_i), \  \ \vv_i=\frac{1}{m_i}(m\vv)_i
\end{equation}
where $\xx_p$ and $\xx_i$ are particle and grid node locations and $N(\xx_p-\xx_i)$
represent interpolating functions defined on the grid. After a physical update of the
momentum is done on the grid, new grid velocities $\tilde{\vv}_i$ are then directly
interpolated to particles as
\begin{equation}
\tilde{\vv}_p=\sum_i\tilde{\vv}_iN(\xx_p-\xx_i).
\end{equation}
With this simple convention, linear and angular momentum are conserved in the transfer
from particle to grid as long as the interpolating functions satisfy a partition of unity
property. In the transfer from grid to particle, linear momentum is conserved, but
angular momentum is not. Notably, these transfers are linear operations, and
since there are typically many more particle than grid degrees of freedom, there are
particle velocity null modes that are lost when transferring to grid. Since the PIC
transfer from grid to particle is just interpolation, this process can be seen as
filtering out particle velocity modes that are not seen on the grid. The loss of the
kinetic energy in these modes is what leads to the excessive dissipation of PIC.

The energy loss in PIC style transfers is unacceptable for many application areas, and
FLIP style transfers can be used instead. FLIP uses the same transfer from particles to
grid as PIC, however with FLIP, velocities are incremented by interpolated differences in
grid velocities (rather than directly interpolated as in PIC) when transferring from grid
to particles
\begin{equation}
\tilde{\vv}_p=\vv_p+\sum_i(\tilde{\vv}_i-{\vv}_i)N(\xx_p-\xx_i).
\end{equation}
Since velocities are incremented, rather than overwritten with information from the grid,
energy in particle null modes is not lost and thus the excessive dissipation is
avoided. However, these modes are still invisible to the grid, since the transfer from
particle to grid is the same in PIC and FLIP. Thus, although these modes are not lost,
they have no direct effect on the governing physics which can lead to unpredictable
behavior like those discussed in
\cite{langdon:1970:finite_grid,okuda:1972:PIC_inst,brackbill:1988:ringing-pic,gritton:2014:ringing,brackbill:1998:magnetohydrodynamics,bardenhagen:2004:generalized-mpm}.

We present a new technique designed to retain the filtering property of the original PIC
transfers to guarantee stable behavior. We show that by representing particle velocities
as locally affine, rather than locally constant, particle/grid transfers can be defined
that: (1) filter out null modes invisible to the grid, (2) have dissipation comparable to
that of FLIP and (3) conserve angular and linear momentum (both from particle to grid and
grid to particle). Furthermore, this is all done with simple mass lumping foregoing the
need for poorly conditioned full mass matrices.

There are a few existing approaches that use similar ideas to what we propose. Our work
builds on that of Jiang et al \cite{jiang:2015:apic}. The transfers used there are
discretely angular momentum conserving only for explicit symplectic Euler integration. For
any other integration scheme, angular momentum may be gradually lost. Wallstedt and
Guilkey also augment particles with derivatives of the field variables from grid to reduce
dissipation in \cite{wallstedt:2007:velocity-projection}. However they still use FLIP
style incremental updates and thus still suffer from null mode persistence. Furthermore,
their transfer from grid to particle is not angular momentum conserving. Also, our
approach is similar to some aspects of the Constrained Interpolation Profile (CIP) methods
which also store derivative information to reduce diffusion and improve conservation, but
for semi-Lagrangian interpolation \cite{yabe:2001:CIP}.

\section{Momentum transfers}

The primary difference between our method and a traditional PIC scheme is that particles
represent piecewise affine, rather than constant samples of the velocity field. Thus, in
addition to a sample of the local velocity $\vv_p$, we conceptually represent the velocity
as $\vv(\xx)=\vv_p+\CC_p(\xx-\xx_p)$ local to the particle $\xx_p$. We show that this will
allow us to design a family of transfers that better preserve momentum and energy without
creating persistent null modes on particles.

\subsection{Rigid-Particle-In-Cell (RPIC)}\label{sec:RPIC}

The intuition for our transfers is largely derived from a simpler case: piecewise rigid
body velocity. This can be thought of as defining a velocity field local to $\xx_p$ as
$\vv(\xx)=\vv_p+\CC_p(\xx-\xx_p)$ with skew symmetric $\CC_p$. That is,
$\CC_p=\boldsymbol\omega_p^*$ where $\boldsymbol\omega_p$ is the angular velocity of the
rigid body and $\CC_p=\boldsymbol\omega_p^*$ is the skew symmetric matrix equivalent to
$\CC_p\xx=\boldsymbol\omega_p\times\xx$ for arbitrary vector $\xx$. While this
idealization can also be used with PIC style filtering and both linear and angular
momentum conservation across all transfers, it ultimately suffers from excessive
dissipation similar to PIC \cite{jiang:2015:apic}. Nonetheless, it provides most of the
insights needed for making transfers with general affine conservative so we present them
here.

\subsubsection{Particle to grid}\label{sec:RPICp2g}
With a piecewise rigid assumption, we idealize particle $\xx_p$ as a rigid body consisting
of point masses that the particle distributes to the grid with standard PIC transfer:
$m_{ip}=m_pN(\xx_p-\xx_i)$. That is, rigid body $p$ consists of point masses $m_{ip}$
located at $\xx_i$. Note that this rigid body then has inertia tensor
$\KK_p=\sum_im_{ip}(\xx_i-\xx_p)^*(\xx_i-\xx_p)^{*T}$. Also note that the standard PIC
grid mass is then $m_i=\sum_p m_{ip}$. With this idealization, the linear momentum of the
points in the rigid body are then $(m\vv)_{ip}=m_{ip}(\vv_p+\CC_p(\xx_i-\xx_p))$ where
again $\CC_p$ is assumed to be skew symmetric to represent rigid body velocity. We can
thus define the grid linear momenta to be the sum of the contributions from all rigid
bodies $p$: $(m\vv)_i=\sum_p(m\vv)_{ip}$.

This transfer conserves linear and angular momenta in the following sense. Define the
total linear momentum of all rigid bodies as $\pp^P=\sum_pm_p\vv_p$ and the total angular
momentum (about the origin) as $\ll^P=\sum_p\KK_p\boldsymbol\omega_p+\xx_p\times m_p\vv_p$
(see Section~\ref{sec:RPICproofs} for justification of these definitions). After the
transfer from particle to grid, we have $\pp^G=\sum_i (m\vv)_i$ and $\ll^G=\sum
\xx_i\times (m\vv)_i$ as the analogous quantities defined over the grid. It can be shown
that $\pp^P=\pp^G$ and $\ll^P=\ll^G$ (see Section~\ref{sec:RPICproofs} for details). That
is, we can say that the linear and angular momentum of the grid state is the same as that
of the particle rigid body state after the transfer from particle to grid.

\subsubsection{Grid to particle}\label{sec:RPICg2p}
The transfer from grid to particle is done after a momentum update on the grid. However,
the update of the grid state will give a new $\tilde{\vv}_i$ since the grid node masses
$m_i$ do not change over the step. We design transfers of $\tilde{\vv}_i$ to get
$\tilde{\vv}_p$ and skew $\tilde{\CC}_p$ that give a rigid body state whose linear and
angular momentum are consistent with that of the updated grid state. That is, we want
$\tilde{\vv}_p$ and skew $\tilde{\CC}_p=\tilde{\boldsymbol\omega}_p^*$ such that the new
linear momentum is conserved $\tilde{\pp}^G=\sum_i m_i\tilde{\vv}_i=\sum_p
m_p\tilde{\vv}_p=\tilde{\pp}^P$ and new angular momentum is conserved $\tilde{\ll}^G=\sum
\xx_i\times m_i\tilde{\vv}_i=\sum_p(\KK_p\tilde{\boldsymbol\omega}_p+\xx_p\times
m_p\tilde{\vv}_p)=\tilde{\ll}^P$. If we define the transfer of the linear velocity as with
standard PIC, $\tilde{\vv}_p=\sum_i\tilde{\vv}_iN(\xx_p-\xx_i)$, then linear momentum is
conserved, as with PIC. However, with this transfer alone, angular momentum is
lost. Specifically, it can be shown that local to particle $p$,
$\tilde{\ll}_p=\sum_i(\xx_i-\xx_p)\times m_{ip}\tilde{\vv}_i$ is lost. This arises from
representing the information in the grid state $m_{ip}\tilde{\vv}_i$ as only
$m_p\vv_p$. Clearly, one particle can not represent the angular momentum seen on the grid
in $m_{ip}\tilde{\vv}_i$. The idea is to represent that angular momentum in a rigid body,
rather than a simply translating body to prevent the loss. Thus, if we define angular
velocity $\tilde{\boldsymbol\omega}_p$ to be
$\tilde{\boldsymbol\omega}_p=\KK_p^{-1}\tilde{\ll}_p$ (and
$\tilde{\CC}_p=\tilde{\boldsymbol\omega}_p^*$), then a simple argument shows that both
linear and angular momentum are conserved in the transfer from grid to particle. That is,
the transfers give a rigid body state whose linear and angular momentum are consistent
with that of the updated grid state. See Section~\ref{sec:RPICproofs} for proofs of these
claims.

\subsection{Affine-Particle-In-Cell (APIC)}

For APIC, we will extend the particle-wise, local velocity field to be an arbitrary affine
function as $\vv(\xx)=\vv_p+\CC_p(\xx-\xx_p)$. Here the matrix $\CC_p$ is fully arbitrary,
unlike the skew symmetric view in RPIC. The problem then is to determine the transfers
from particle to grid and vice versa. This can be done in a manner directly analogous to
what was presented in Section~\ref{sec:RPIC}, and we provide those details in
Section~\ref{sec:unified}. However, when developing a scheme that is perfectly
conservative over the entire time step (i.e., both transfers and grid updates are
conservative), a more general notion of transfer is useful. The discussion of transfers so
far has assumed that information will be transferred from particles to the grid and then
immediately back to particles without any other changes in grid or particle positions.
While we show that these transfers can be made perfectly conservative, this is typically
not enough in practice. The point of hybrid particle/grid schemes is that part of the
evolution will occur on the grid. This introduces an element of time into the conservation
problem.  For example, immediately following the transfer from particle to grid, the angular
momentum should be computed as  $\ll^G = \sum_i \xx_i^n\times m_i^n\vv_i^n$.  Before the
transfer back to particles, the grid state will have changed, and angular momentum will be
computed as $\ll^G = \sum_i \xx_i^{n+1}\times m_i^n\vv_i^{n+1}$. We introduce a degree of
flexibility into the definition of the APIC transfers to account for this.  When
transferring to the particles, we have access to $\xx_p^n$, $\xx_i^n$, $\xx_p^{n+1}$, and
$\xx_i^{n+1}$, which gives us more possible options.  We are also free to choose the state
that we store.  For example with RPIC, rather than storing angular velocity $\op$ as
state, we could store rotational angular momentum $\tilde{\ll}_p$.  This additional
flexibility is very useful, since it allows us to obtain additional properties from the
method.  We require our transfers to be generally of the form described above, subject to
the additional flexibility that has been noted.

Now that we have broadened our search space of possible transfers, we need to narrow down
the possibilities.  We narrow the field of choices down to a single scheme by enforcing
three properties:
\begin{enumerate}\label{eq:APIC_principles}
\item A globally affine velocity field should be preserved across transfers from particles
  to the grid and back when moving particles and moving grids are ignored (for example
  when $\dt = 0$).\label{it:affine}
\item The transfers should conserve linear and angular momentum, even when the
  complications of grid-based evolution, moving grids, and moving particles are taken into
  account.\label{it:conserve}
\item A simulation with a single particle is stable but non-dissipative when moving grids
  and moving particles are taken to account but additional grid-based influences (forces,
  etc.) are ignored.\label{it:stable}
\end{enumerate}
Property~\ref{it:affine} is what it means to be an APIC scheme; it is a PIC-style transfer
that preserves affine velocity fields.  Note that this property should only be enforced
under very strict circumstances ($\dt = 0$), since affine velocity fields should be able
to change due to advection.  Property~\ref{it:conserve} ensures that the entire scheme
will conserve linear and angular momentum provided that the grid-based scheme also
conserves these quantities.

Property~\ref{it:stable} is a non-obvious but crucial requirement.
The other properties do not uniquely determine a transfer; they only narrow it down to a
one-parameter family of transfers.  These transfers tend to behave similarly
\textit{except} when one particle moves far enough from other particles that it is able to
evolve in isolation.  For one particular member of this family, a lone particle will
evolve by not changing.  For the rest of the members of this family, part of the
particle's state tends to explode or decay exponentially when the particle evolves in
isolation.  Exponential decay is not desirable, and exponential growth is intolerable.
This leads us to choose the stability criterion to narrow the possibilities down to one
set of transfers.  We present these transfers in the context of the MPM method in which we
use them in Sections~\ref{sec:p2g} and~\ref{sec:g2p}. Also, we present a derivation of the
transfers from the properties 1-3 in Section~\ref{sec:APIC_derivation}.

\section{Method}

We demonstrate the behavior of our transfers on MPM simulations of hyperelasticity. Here
we outline the governing equations and establish some notation used throughout the
exposition.

\subsection{Equations}

Let $\xx = \ph(\XX,t)$ be the mapping from material coordinates $\XX$ to world coordinates
$\xx$.  Let $\VV$ and $\vv$ be the Lagrangian and Eulerian velocities.  $\FF$ is the
deformation gradient, and $J$ is its determinant.  That is,
\begin{align}
  \VV(\XX,t) &= \px{\xx}{t}(\XX,t) \\
  \vv(\xx,t) &= \VV(\ph^{-1}(\xx,t),t) \\
  \FF(\XX,t) &= \px{\xx}{\XX}(\XX,t) \\
  J &= \det(\FF)
\end{align}
With these definitions, the evolution equations are
\begin{align}
  \rho \frac{D\vv}{Dt} &= \nabla\cdot\sg,
\end{align}
where the Cauchy stress $\sg$ is related to the first Piola-Kirchhoff stress $\PP$ and
hyperelastic energy density $\Psi$ through
\begin{align}
  \sg &= \frac{1}{J} \PP \FF^T \\
  \PP &= \px{\Psi}{\FF}
\end{align}
The state of stress in hyperelastic materials is simply related to $\FF$ as $\Psi(\FF)$
and $\PP(\FF)$ where the total internal potential energy $\Phi$ is
\begin{align}
  \Phi(t) &= \int_{\Omega_0} \Psi(\FF(\XX,t))\,d\XX.
\end{align}
Since we will not have access to a reference configuration, we must evolve our deformation
gradient according to
\begin{align}
  \px{\FF}{t}(\XX,t) &= \px{\vv}{\xx}(\ph(\XX,t),t) \FF(\XX,t).
\end{align}
We seek to conserve total momentum $\pp(t)$ and total angular momentum $\ll(t)$, which are given by
\begin{align}
  \pp(t) &= \int_{\Omega} \rho(\xx,t) \vv(\xx,t)\,d\xx \\
  \ll(t) &= \int_{\Omega} \xx \times \rho(\xx,t) \vv(\xx,t)\,d\xx
\end{align}
For completeness, kinetic energy is
\begin{align}
  T(t) &= \int_{\Omega} \rho(\xx,t) \|\vv(\xx,t)\|^2 \,d\xx,
\end{align}
and total energy is $E = T + \Phi$.

\subsection{Transfer to grid}\label{sec:p2g}

Each particle $\xx_p^n$ stores mass $m_p$, velocity $\vv_p^n$, and the additional matrix
$\BB_p^n$. As we are using MPM, we also store a deformation gradient $\FF_p^n$ on
particles. Note that particle masses $m_p$ do not have a time superscript because they are
constant (and thus never updated from the grid) to account for conservation of mass. We
first use our weights to interpolate mass and momentum to the grid.
\begin{align}
m_i^n &= \sum_p m_p w_{ip}^n \\
\DD_p^n &= \sum_i w_{ip}^n (\xx_i^n-\xx_p^n) (\xx_i^n-\xx_p^n)^T \\
m_i^n \vv_i^n &= \sum_p w_{ip}^n m_p (\vv_p^n + \BB_p^n (\DD_p^n)^{-1} (\xx_i^n-\xx_p^n))
\end{align}
The velocity $\vv_i^n$ is obtained by division.  Note that unlike with $m_p$, we specify a
time superscript on grid mass $m_i^n$, since it will change each time step. The additional
matrix $\DD_p^n$ used in the transfer is similar to an inertia tensor (but for an affine
rather than rigid motion). Similarly, $\BB_p^n$ contains angular momentum information and
the local affine velocity field is conceptually $\vv_p^n + \BB_p^n (\DD_p^n)^{-1}
(\xx_i^n-\xx_p^n)$ with matrix $\CC_p^n=\BB_p^n (\DD_p^n)^{-1}$. We will elaborate on
these properties later when we prove conservation.

\subsection{Grid evolution}

At this point, we have transferred state from particle to grid, and we are ready to apply
forces and perform our grid-based evolution.  We must update grid velocity $\vt_i^{n+1}$,
position $\xt_i^{n+1}$, and deformation gradient $\FF_p^{n+1}$.  The update of grid
positions to $\xt_i^{n+1}$ is purely conceptual.  Our implementation uses fixed Cartesian
grids.

An important aspect of allowing for exact conservation of linear and angular momentum
during particle/grid transfers is that conservation of the entire method can be achieved
by combining with one of the many conservative integrators used for updating the grid
state
\cite{gonzalez:2000:exact-conserving,laursen:2001:energy-conserving,simo:1992:energy-momentum,simo:1992:symplectic,kane:1999:variational,lew:2004:variational}. We
introduce a parameter $\lambda$, which allows us to consider an entire family of methods
that conserve linear and angular momentum.  This family contains two notable members:
symplectic Euler ($\lambda = 0$) and midpoint rule ($\lambda = \frac{1}{2}$).  The schemes
$\lambda = 0$ and $\lambda = 1$ are both explicit; the rest are implicit.  We use midpoint
rule for all of our examples.  Note that schemes such as forward Euler, backward Euler,
and trapezoid rule do not conserve angular momentum and thus are not suitable for our
purposes.  Our family of grid-based updates is
\begin{align}
\FF_p^{n+1} &= \gp{\II+\sum_i(\xt_i^{n+1} - \xx_i^n)(\nabla w_{ip}^n)^T}\FF_p^n \\
\FF_p^{n+\lambda} &= (1-\lambda) \FF_p^n + \lambda \FF_p^{n+1} \\
\vt_i^{n+1} &= \vv_i^n + \frac{\dt}{m_i^n}\ff_i^{n+\lambda} \\
\xt_i^{n+1} &= \xx_i^n + \dt(\lambda \vv_i^n + (1-\lambda) \vt_i^{n+1})
\end{align}
The velocity update rule uses forces $\ff_i^{n+\lambda}$, which we define from a potential
energy function $\Phi^{n+\lambda}$, which we compute from an energy density
$\Psi_p(\FF_p)$.  Our rules for computing potential energy $\Phi^{n+\lambda}$, force
$\ff_i^{n+\lambda}$, and product by force derivatives are
\begin{align}
\Psi_p^{n+\lambda} &= \Psi_p\gp{\FF_p^{n+\lambda}} \\
\Phi^{n+\lambda} &= \sum_p V_p \Psi_p^{n+\lambda} \\
\PP_p^{n+\lambda} &= \PP_p\gp{\FF_p^{n+\lambda}} \\
\ff_i^{n+\lambda} &= \sum_p V_p \PP_p^{n+\lambda} (\FF_p^n)^T \nabla w_{ip}^n \\
\AA_p &= \px{\PP_p}{\FF_p} : \gp{\sum_i \dv_i (\nabla w_{ip}^n)^T \FF_p^n} \label{eqn:force-derivative} \\
\sum_j \gp{\px{\ff_i}{\xx_j}} \dv_j &= \sum_p V_p \AA_p (\FF_p^n)^T \nabla w_{ip}^n \label{eqn:force-derivative-helper}
\end{align}
Here, $\PP_p$ is the first Piola-Kirchhoff stress tensor.  Defining forces through an
energy ensures angular momentum conservation; the particular constitutive model does not
matter.

\subsection{Transfer to particles}\label{sec:g2p}

With grid evolution completed, we have updated grid locations $\xt_i^{n+1}$ and velocities
$\vt_i^{n+1}$.  What remains is to transfer this information back to particles.  We do
this using the transfers
\begin{align}
\vv_p^{n+1} &= \sum_i w_{ip}^n \vt_i^{n+1} \\
\BB_p^{n+1} &= \frac{1}{2}\sum_i w_{ip}^n \gp{\vt_i^{n+1} (\xx_i^n-\xx_p^n+\xt_i^{n+1}-\xx_p^{n+1})^T + (\xx_i^n-\xx_p^n-\xt_i^{n+1}+\xx_p^{n+1})(\vt_i^{n+1})^T} \label{eqn:update-bb} \\
\xx_p^{n+1} &= \sum_i w_{ip}^n \xt_i^{n+1} \\
\FF_p^{n+1} &= \gp{\II+\sum_i(\xt_i^{n+1} - \xx_i^n)(\nabla w_{ip}^n)^T}\FF_p^n
\end{align}
This completes the specification of our angular-momentum-conserving family of APIC
schemes.

\subsection{Interpolation weights}

As with PIC, we use weights to transfer information between the two representations.
While the choice of weights is flexible, we require them to satisfy some important
properties.  Let $N(\xx)$ be an interpolation kernel, which must be chosen to satisfy
\begin{align}
\sum_i N(\xx_p^n - \xx_i^n) &= 1 \\
\sum_i \xx_i^n N(\xx - \xx_i^n) &= \xx
\end{align}
for any $\xx$.  The kernel $N(\xx)$ is used to define interpolation weights and weight
gradients as $w_{ip}^n = N(\xx_p^n - \xx_i^n)$ and $\nabla w_{ip}^n = \nabla N(\xx_p^n -
\xx_i^n)$.  The properties above lead to properties for $w_{ip}^n$ and $\nabla w_{ip}^n$.
\begin{align}
\sum_i w_{ip} &= 1 \\
\sum_i w_{ip} \xx_i^n &= \xx_p^n \\
\sum_i w_{ip} (\xx_i^n-\xx_p^n) &= \z \\
\sum_i \xx_i^n (\nabla w_{ip}^n)^T &= \II
\end{align}
With these weights defined, we can start describing the method.

\section{Implementation details}

\subsection{Implicit midpoint as minimization problem}\label{sec:optimization}

The grid update is in general implicit, including the midpoint rule ($\lambda =
\frac{1}{2}$).  Since this is the member that we implemented and recommend using, we
restrict our attention here to this case.  We also demonstrate symplectic Euler and
backward Euler as grid update schemes for comparison in some of our numerical experiments.
Symplectic Euler is explicit and does not require the optimization treatment that follows.
Backward Euler is not a member of the family described in this paper; we compare against
it for reference.

We solve the resulting nonlinear systems of equations following an optimization-stabilized
Newton-Raphson solver framework \cite{gast:2014:sca,gast:2015:tvcg}.  The implicit
midpoint scheme for MPM grid nodes is
\begin{align*}
  \xt_i^{n+1} &= \xx_i^n + \dt \gp{\frac{\vv_i^n+\vt_i^{n+1}}{2}}, \\
  \vt_i^{n+1} &= \vv_i^n + \frac{\dt}{m_i^n}\ff_i\gp{\frac{\xx_i^n+\xt_i^{n+1}}{2}}.
\end{align*}
Eliminating $\xt_i^{n+1}$ gives
\begin{align*}
  m_i^n\frac{\vt_i^{n+1}-\vv_i^n}{\dt} &= \ff_i\gp{\frac{\xx_i^n+\xx_i^n+\dt\gp{\frac{\vv_i^n+\vt_i^{n+1}}{2}}}{2}} \\
  &= \ff_i\gp{\xx_i^n+\frac{\dt}{4}(\vv_i^n+\vt_i^{n+1})}.
\end{align*}
Changing to the variable $\dv_i=\vt_i^{n+1}-\vv_i^n$,
\begin{align}\label{eqn:system-dv}
  m_i^n\dv_i &= \dt\ff_i\gp{\xx_i^n+\frac{\dt}{2}\vv_i^n+\frac{\dt}{4}\dv_i}.
\end{align}
The corresponding minimization objective function is
\begin{align*}
  E(\dv_i)&=\sum_i \frac{m_i^n}{8} \norm{\dv_i}^2 + \Phi\gp{\xx_i^n+\frac{\dt}{2}\vv_i^n+\frac{\dt}{4}\dv_i}.
\end{align*}
This is similar to the corresponding objective for backward Euler, which is
\begin{align*}
  E_{be}(\dv_i)&=\sum_i\frac{m_i^n}{2} \norm{\dv_i}^2 + \Phi\gp{\xx_i^n+\dt\vv_i^n+\dt^2\dv_i}.
\end{align*}
The minimum of $E$ occurs when
\begin{align*}
  \gg_i &= \px{E}{\dv_i} \\
  &= \px{}{\dv_i}\gp{\sum_j \frac{m_j}{8} \norm{\dv_j}^2 + \Phi\gp{\xx_j^n+\frac{\dt}{2}\vv_j^n+\frac{\dt}{4}\dv_j}} \\
  &= \frac{m_i^n}{4} \dv_i - \frac{\dt}{4}\ff_i\gp{\xx_i^n+\frac{\dt}{2}\vv_i^n+\frac{\dt}{4}\dv_i},
\end{align*}
Note that $\gg(\dv_i) = \z$ is just \eqref{eqn:system-dv}, so minimizing $E$ is equivalent
to solving \eqref{eqn:system-dv}.  Multiplying the derivative of $\gg$ by some vector
$\delta\uu_i$ will be necessary.
\begin{align*}
  \sum_j \px{\gg_i}{\dv_j} \delta\uu_j
  &= \sum_j \px{}{\dv_j}\gp{\frac{m_i^n}{4} \dv_i - \frac{\dt}{4}\ff_i\gp{\xx_i^n+\frac{\dt}{2}\vv_i^n+\frac{\dt}{4}\dv_i}} \delta\uu_j \\
  &= \frac{m_i^n}{4} \delta\uu_i - \frac{\dt^2}{16}\sum_j \px{\ff_i}{\xx_j}\gp{\xx_i^n+\frac{\dt}{2}\vv_i^n+\frac{\dt}{4}\dv_i} \delta\uu_j
\end{align*}
This in turn requires a matrix-vector multiply by the force derivative, which is done
using \eqref{eqn:force-derivative} and \eqref{eqn:force-derivative-helper}.

\subsection{Momentum conservation on incomplete convergence}

The conservation properties of our method (see
Section~\ref{sec:proof-conservation-momentum}) depend on solving \eqref{eqn:system-dv} to
convergence.  If this is not done, conservation will be only approximate.  We note,
however, that this is not a fundamental problem.  One way to track down the source of the
problem is to label every vector a \textit{velocity-like} or \textit{force-like}.  Assume
initial velocity is zero and all forces are momentum-conserving.  Then, we can note some
rules about how these types of vector should behave:
\begin{enumerate}
\item A force-like vector will sum to zero.
\item A velocity-like vector will sum to zero when scaled by mass.
\item Scaling a velocity-like vector by mass produces a force-like vector.
\item Scaling a force-like vector by inverse mass produces a velocity-like vector.
\item Scaling a vector by a constant preserves its type.
\item Adding vectors is only permitted if they have the same type; the type is preserved.
\item In the matrix-vector multiply $\delta\ff_i = \sum_j \px{\ff_i}{\xx_j} \delta\uu_j$,
  $\delta\uu_j$ must be velocity-like, and $\delta\ff_i$ will be force-like.
\item Dot product is only allowed if one vector is force-like and the other is
  velocity-like.  (This is done, for example, when computing kinetic energy.)
\end{enumerate}
As long as these rules are followed, the velocity will be velocity-like, which implies
conservation of linear momentum (the last rule is not strictly required, but we can
enforce it anyway).  Propagating these labels through the algorithm (Newton's method, line
searches, conjugate gradient, etc.) is straightforward and breaks down only inside the
conjugate gradient solver.  The source of the problem is that $\pp$, $\rr$, and $\ss$ must
be of the same type (see Algorithm~\ref{alg:cg}), so that $\ss \gets \AA \pp$ means the
operator $\AA$ must take and produce the same type of vector.  The system we are solving
takes the general form
\begin{align*}
  \AA_1 \delta \vv &= \delta \ff \qquad \AA_1 = \MM + \zeta \px{\ff}{\xx},
\end{align*}
where $\MM$ is a diagonal mass matrix, $\zeta$ is a scalar, $\delta \vv$ is a
velocity-like vector, $\delta \ff$ is a force-like vector.  The operator $\AA_1$ takes
velocity-like vectors and produces force-like vectors, which is a problem.  We can avoid
that problem by rewriting
\begin{align*}
  \AA_2 \delta \vv &= \MM^{-1} \delta \ff \qquad \AA_1 = \II + \MM^{-1}\zeta \px{\ff}{\xx}.
\end{align*}
Now, $\AA_2$ takes velocity-like vectors and returns velocity-like vectors.
Unfortunately, this $\AA_2$ is not symmetric.

The conjugate gradient operates on vectors in only a few ways: matrix-vector multiply,
vector operations, and inner product.  Note that the inner product used does not need to
be the standard inner product:  $\AA$ is only required to be symmetric with respect
to the inner product chosen.  That is, $\langle \AA \uu, \vv \rangle = \langle \uu, \AA
\vv \rangle$ for any $\uu$ and $\vv$.  Note that $\AA_2$ is symmetric with respect to the
mass inner product $\langle \uu, \vv \rangle = \uu^T \MM \vv$.  Using this modified system
and a mass inner product for conjugate gradient is a perfectly acceptable means of solving
the linear system.  Furthermore, all vectors in the conjugate gradient algorithm are now
velocity-like, which allows us to label all of our vectors.  This in turn guarantees
conservation of momentum, even if our solver is not fully converged.

\begin{algorithm}[t]
\caption{Conjugate Gradient}
\label{alg:cg}
\begin{algorithmic}[1]
\Procedure{Conjugate\textendash Gradient}{$\AA$, $\xx$, $\bb$}
\State $\rr \gets \bb - \AA \xx$
\State $\pp \gets \rr$
\State $\gamma \gets \langle \rr, \rr \rangle$
\While{not converged}
\State $\ss \gets \AA \pp$
\State $\displaystyle \alpha \gets \frac{\gamma}{\langle \pp, \ss\rangle}$
\State $\xx \gets \xx + \alpha \pp$
\State $\rr \gets \rr - \alpha \ss$ \Comment{$\rr$ and $\ss$ have the same type}
\State $\kappa \gets \langle \rr, \rr\rangle$
\State $\beta \gets \frac{\kappa}{\gamma}$
\State $\pp \gets \rr + \beta \pp$ \Comment{$\rr$ and $\pp$ have the same type}
\State $\gamma \gets \kappa$
\EndWhile
\EndProcedure
\end{algorithmic}
\end{algorithm}

\subsection{CFL condition}

We choose our time step size $\dt$ so that no particle will travel more than the grid
spacing $\dx$ in one time step.  We approximate this by assuming that these particles
travel with the initial grid velocity $\vv_i^n$.  While this does not take into account
the potentially dramatic affect of forces, we note that our method is implicit and can
tolerate such errors.
\begin{align*}
m_i^n \vv_i^n &= \sum_p w_{ip}^n m_p (\vv_p^n + \BB_p^n (\DD_p^n)^{-1} (\xx_i^n-\xx_p^n)) \\
m_i^n \|\vv_i^n\| &= \norm{\sum_p w_{ip}^n m_p (\vv_p^n + \BB_p^n (\DD_p^n)^{-1} (\xx_i^n-\xx_p^n))} \\
&\le \norm{\sum_p w_{ip}^n m_p \vv_p^n} + \norm{\sum_p w_{ip}^n m_p\BB_p^n (\DD_p^n)^{-1} (\xx_i^n-\xx_p^n)} \\
&\le \sum_p w_{ip}^n m_p \|\vv_p^n\| + \sum_p w_{ip}^n m_p \|\BB_p^n (\DD_p^n)^{-1} (\xx_i^n-\xx_p^n)\| \\
&\le \sum_p w_{ip}^n m_p \|\vv_p^n\| + \sum_p w_{ip}^n m_p \|\BB_p^n\|_F \|(\DD_p^n)^{-1} (\xx_i^n-\xx_p^n)\|
\end{align*}
Interpolation stencil support is bounded by $\|\xx_i^n-\xx_p^n\| \le \kappa \dx$.  If we
also assume $D_p = k \II$, then $\|(\DD_p^n)^{-1} (\xx_i^n-\xx_p^n)\| \le \frac{\kappa}{k}
\dx$.
\begin{align*}
m_i^n \|\vv_i^n\| &\le \sum_p w_{ip}^n m_p \|\vv_p^n\| + \sum_p w_{ip}^n m_p \|\BB_p^n\|_F \|(\DD_p^n)^{-1} (\xx_i^n-\xx_p^n)\| \\
&\le \sum_p w_{ip}^n m_p \gp{\|\vv_p^n\| + \frac{\kappa}{k} \dx \|\BB_p^n\|_F} \\
&\le \gp{\sum_p w_{ip}^n m_p} \max_p\gp{\|\vv_p^n\| + \frac{\kappa}{k} \dx \|\BB_p^n\|_F} \\
&= m_i^n \max_p\gp{\|\vv_p^n\| + \frac{\kappa}{k} \dx \|\BB_p^n\|_F} \\
\|\vv_i^n\| &\le \max_p\gp{\|\vv_p^n\| + \frac{\kappa}{k} \dx \|\BB_p^n\|_F}
\end{align*}
In the case of both quadratic and cubic interpolation, $\frac{\kappa}{k} \dx =
\frac{6\sqrt{d}}{\dx}$, where $d$ is the dimension.  A reasonable CFL condition is then
\begin{align*}
\dt &\le \frac{\nu \dx}{\max_p\gp{\|\vv_p^n\| + \frac{\kappa}{k} \dx \|\BB_p^n\|_F}}.
\end{align*}
We use $\nu = 1$ for our examples.

\section{Notes and analysis}

Here we discuss a number of aspects and useful properties of the schemes we have proposed.

\subsection{RPIC transfer properties}\label{sec:RPICproofs}
The RPIC transfers outlined in Sections~\ref{sec:RPICp2g} from particle momenta to grid
momenta are
\begin{equation}
(m\vv)_i=\sum_pm_pN(\xx_p-\xx_i)(\vv_p+\CC_p(\xx_i-\xx_p)).
\end{equation}
The transfers from updated grid velocities $\tilde{\vv}_i$ to new particles velocities
$\tilde{\vv}_p$ and angular velocities $\tilde{\boldsymbol\omega}_p$ as outlined in
Section~\ref{sec:RPICg2p} are
\begin{equation}\label{eq:RPICg2p}
\tilde{\vv}_p=\sum_i N(\xx_p-\xx_i)\tilde{\vv_i}, \ \ \tilde{\boldsymbol\omega}_p=\KK_p^{-1}\left(\sum_i(\xx_i-\xx_p)\times m_pN(\xx_p-\xx_i)\tilde{\vv}_i\right).
\end{equation}
These transfers conserves total linear and angular momenta. To define the total linear and
angular momenta of the particles, we think of them as a collection of rigid bodies, each
made up of individual point masses $m_{ip}=m_pN(\xx_p-\xx_i)$ located at the grid nodes
$\xx_i$. Then the total momenta of the collection of rigid bodies is the sum of the
contributions from each respective point mass. That is, the total linear momentum $\pp^P$
of the particles is
\begin{equation}
\pp^P=\sum_p\sum_i m_{ip}\left(\vv_p+\CC_p(\xx_i-\xx_p)\right)
\end{equation}
and the total angular momentum (computed about the origin) is
\begin{equation}
\ll^P=\sum_p\sum_i \xx_i \times m_{ip}\left(\vv_p+\CC_p(\xx_i-\xx_p)\right).
\end{equation}
These quantities are defined in a more obvious manner on the grid as
\begin{equation}
\pp^G=\sum_i m_i\vv_i \ \ \textrm{and} \ \ \ll^G=\sum_i \xx_i \times m_i\vv_i.
\end{equation}
In the following, we will show that after the transfer from particle to grid,
$\pp^P=\pp^G$ and $\ll^P=\ll^G$ and after the transfer from grid to particle
$\tilde{\pp}^P=\tilde{\pp}^G$ and $\tilde{\ll}^P=\tilde{\ll}^G$

\subsubsection{Particle to grid: conservation of linear momentum}
The total linear momenta are equal after the transfer, which can be seen simply from
\begin{equation}
\begin{split}
\pp^P & =\sum_p\sum_im_pN(\xx_p-\xx_i)\left(\vv_p+\CC_p(\xx_i-\xx_p)\right) \\
& =\sum_i\sum_p m_pN(\xx_p-\xx_i)\left(\vv_p+\CC_p(\xx_i-\xx_p)\right)=\sum_i m_i\vv_i=\pp^G.
\end{split}
\end{equation}
However, it is also useful to note that
\begin{equation}\label{eq:Rp}
\pp^P=\sum_p\sum_i m_pN(\xx_p-\xx_i)\left(\vv_p+\CC_p(\xx_i-\xx_p)\right)=\sum_pm_p\vv_p.
\end{equation}
That is, the $\CC_p$ contribute no net linear momentum. This holds since $\sum_p\sum_i
m_pN(\xx_p-\xx_i)\vv_p=\sum_pm_p\vv_p$ where we assume a partition of unity property of
the grid interpolating functions $\sum_i N(\xx-\xx_i)=1$ and
$\sum_i\sum_pm_pN_i(\xx_p)\CC_p(\xx_i-\xx_p)=\mb{0}$. This can be seen from
\begin{equation}
\sum_i\sum_pm_pN(\xx_p-\xx_i)\CC_p\xx_p=\sum_pm_p\CC_p\xx_p\sum_iN(\xx_p-\xx_i)=\sum_pm_p\CC_p\xx_p
\end{equation}
again using partition of unity and lastly
\begin{equation}
\sum_i\sum_pm_pN(\xx_p-\xx_i)\CC_p\xx_i=\sum_pm_p\CC_p\sum_iN(\xx_p-\xx_i)\xx_i=\sum_pm_p\CC_p\xx_p
\end{equation}
where we assume that the grid interpolation function exactly interpolate linear functions,
which leads to $\sum_iN(\xx-\xx_i)\xx_i=\xx$.

\subsubsection{Particle to grid: conservation of angular momentum}
The transfer also conserves total angular momentum since
\begin{equation}
\begin{split}
\ll^P & = \sum_p\sum_i\xx_i\times m_pN(\xx_p-\xx_i)\left(\vv_p+\CC_p(\xx_i-\xx_p)\right) \\
& = \sum_i\xx_i\times \sum_p m_pN(\xx_p-\xx_i)\left(\vv_p+\CC_p(\xx_i-\xx_p)\right)=\sum_i\xx_i\times m_i\vv_i=\ll^G.
\end{split}
\end{equation}
Also, this formula can be expressed as
\begin{equation}\label{eq:Rl}
\ll^P=\sum_p\xx_p\times m_p\vv_p+\sum_l \KK_p\boldsymbol\omega_p.
\end{equation}
The first term $\sum_p\xx_p\times m_p\vv_p=\sum_p\sum_i\xx_i\times m_pN(\xx_p-\xx_i)\vv_p$
can be seen from partition of unity. The second term is clear when noting
\begin{equation}
\sum_i\xx_p\times m_pN(\xx_p-\xx_i)\CC_p(\xx_i-\xx_p)=\xx_p\times m_p\CC_p(\sum_iN(\xx_p-\xx_i)\xx_i-\xx_p)=\mb{0}
\end{equation}
since $\sum_iN(\xx_p-\xx_i)\xx_i=\xx_p$ and $\sum_iN(\xx_p-\xx_i)=1$, combined with
\begin{equation}
\sum_i \xx_i\times m_pN(\xx_p-\xx_i)\CC_p(\xx_i-\xx_p)=\sum_im_pN(\xx_i-\xx_p)\xx_i^*(\xx_i-\xx_p)^{*^T}\boldsymbol\omega_p
\end{equation}
yields
\begin{equation}
\sum_i \xx_i\times m_pN(\xx_p-\xx_i)\CC_p(\xx_i-\xx_p)=\sum_im_pN(\xx_i-\xx_p)(\xx_i-\xx_p)^*(\xx_i-\xx_p)^{*^T}\boldsymbol\omega_p=\KK_p\boldsymbol\omega_p.
\end{equation}
Recall we use $\xx^*$ to denote the matrix that expresses $\xx \times \yy=\xx^*\yy$ and
$\CC_p=\boldsymbol\omega^*$. This result says that the total angular momentum of the
particles (about the origin) is equal to the sum of the contributions from the conceptual
rigid body centers of mass and the contribution from the angular momentum at each
particle.

\subsubsection{Grid to particle: conservation of linear momentum}
The transfers in Equation~\ref{eq:RPICg2p} conserve linear momentum since
\begin{equation}
\begin{split}
\tilde{\pp}^P & =\sum_pm_p\tilde{\vv}_p=\sum_pm_p\sum_iN(\xx_p-\xx_i)\tilde{\vv}_i \\
& = \sum_i\sum_pm_pN(\xx_p-\xx_i)\tilde{\vv}_i=\sum_im_i\tilde{\vv}_i=\tilde{\pp}^G
\end{split}
\end{equation}
where we use the expression for $\tilde{\pp}^P$ derived in Equation~\ref{eq:Rp}.

\subsubsection{Grid to particle: conservation of angular momentum}
The transfers in Equation~\ref{eq:RPICg2p} also conserve angular momentum. We can show
this using Equations~\ref{eq:Rl} and Equation~\ref{eq:RPICg2p} to express the new total
particle angular momentum as
\begin{equation}\begin{split}
\tilde{\ll}^P & =\sum_p \xx_p\times m_p\tilde{\vv}_p+\sum_p\KK_p\tilde{\boldsymbol\omega}_p=\sum_p \xx_p\times m_p\sum_iN(\xx_p-\xx_i)\tilde{\vv}_i+\sum_p\sum_i(\xx_i-\xx_p)\times m_pN(\xx_p-\xx_i)\tilde{\vv}_i \\
& = \sum_i\xx_i\times \sum_pm_pN(\xx_p-\xx_i)\tilde{\vv}_i=\sum_i\xx_i\times m_i\tilde{\vv}_i=\tilde{\ll}^G
\end{split}
\end{equation}

\subsection{APIC conservation of linear momentum}\label{sec:proof-conservation-momentum}

The APIC scheme is naturally divided into three steps; we show that each step
independently conserves linear momentum.  The first step is the transfer of information
from particle to grid.  We see that the initial particle momentum $\pp^{P,n}$ is
equal to the grid momentum after the transfer $\pp^{G,n}$.
\begin{align*}
\pp^{G,n} &= \sum_i m_i^n \vv_i^n \\
&= \sum_i \sum_p w_{ip}^n m_p (\vv_p^n + \BB_p^n (\DD_p^n)^{-1} (\xx_i^n-\xx_p^n)) \\
&= \sum_i \sum_p w_{ip}^n m_p \vv_p^n + \sum_i \sum_p w_{ip}^n m_p \BB_p^n (\DD_p^n)^{-1} (\xx_i^n-\xx_p^n) \\
&= \sum_p m_p \vv_p^n \sum_i w_{ip}^n + \sum_p m_p \BB_p^n (\DD_p^n)^{-1} \sum_i w_{ip}^n (\xx_i^n-\xx_p^n) \\
&= \sum_p m_p \vv_p^n \\
&= \pp^{P,n}
\end{align*}
Once mass and momentum are on the grid, grid positions and velocities are updated.  We
note that initial grid momentum matches the final grid momentum $\tilde{\pp}^{G,n+1}$.
\begin{align*}
\tilde{\pp}^{G,n+1} &= \sum_i m_i^n \vt_i^{n+1} \\
&= \sum_i m_i^n \gp{\vv_i^n + \frac{\dt}{m_i^n}\ff_i^{n+\lambda}} \\
&= \sum_i m_i^n \vv_i^n + \dt \sum_i \ff_i^{n+\lambda} \\
&= \pp^G + \dt \sum_i \sum_p V_p \PP_p^{n+\lambda} (\FF_p^n)^T \nabla w_{ip}^n \\
&= \pp^G + \dt \sum_p V_p \PP_p^{n+\lambda} (\FF_p^n)^T \sum_i \nabla w_{ip}^n \\
&= \pp^{G,n}
\end{align*}
The final step is transferring information back to particles.  This step is also
conservative since
\begin{align*}
\pp^{P,n+1} &= \sum_p m_p \vv_p^{n+1} \\
&= \sum_p m_p \sum_i w_{ip}^n \vt_i^{n+1} \\
&= \sum_i \vt_i^{n+1} \sum_p m_p w_{ip}^n \\
&= \sum_i m_i^n \vt_i^{n+1} \\
&= \tilde{\pp}^{G,n+1}
\end{align*}
Finally, the entire scheme conserves momentum since $\pp^{P,n+1} = \pp^{P,n}$.

\subsection{APIC conservation of angular momentum}\label{sec:angular-momentum}

We use the permutation tensor in this section.  To make these portions easier to read, we
take the convention that $\AA : \ep$ denotes $A_{\alpha\beta}
\epsilon_{\alpha\beta\gamma}$.  The manipulation $\uu \times \vv = (\vv \uu^T)^T : \ep$ is
used to transition from a cross product into the permutation tensor.

\subsubsection{Transfer to grid}

Our approach to demonstrating angular momentum conservation follows the same three steps.
In this case, we show that $\ll^{P,n} = \ll^{G,n} = \tilde{\ll}^{G,n+1} = \ll^{P,n+1}$,
though the individual steps are more involved.  We begin with the transfer from particles
to the grid.
\begin{align*}
\ll^{G,n} &= \sum_i \xx_i^n \times m_i^n \vv_i^n \\
&= \sum_p\sum_i \xx_i^n \times m_p w_{ip}^n (\vv_p^n + \BB_p^n (\DD_p^n)^{-1} (\xx_i^n-\xx_p^n)) \\
&= \sum_p\sum_i \xx_i^n \times m_p w_{ip}^n \vv_p^n + \sum_p \ll_p^\BB \\
&= \sum_p \xx_p^n \times m_p \vv_p^n + \sum_p m_p (\BB_p^n)^T : \ep \\
&= \ll^{P,n}
\end{align*}
where use has been made from
\begin{align*}
\ll_p^\BB &= \sum_i \xx_i^n \times m_p w_{ip}^n \BB_p^n (\DD_p^n)^{-1} (\xx_i^n-\xx_p^n) \\
&= \sum_i \gp{m_p w_{ip}^n \BB_p^n (\DD_p^n)^{-1} (\xx_i^n-\xx_p^n) (\xx_i^n)^T}^T : \ep \\
&= \gp{m_p \BB_p^n (\DD_p^n)^{-1} \sum_i w_{ip}^n (\xx_i^n-\xx_p^n) (\xx_i^n)^T}^T : \ep \\
&= \gp{m_p \BB_p^n (\DD_p^n)^{-1} \gp{\sum_i w_{ip}^n (\xx_i^n-\xx_p^n) (\xx_i^n-\xx_p^n)^T + \sum_i w_{ip}^n (\xx_i^n-\xx_p^n) (\xx_p^n)^T}}^T : \ep \\
&= \gp{m_p \BB_p^n (\DD_p^n)^{-1} (\DD_p^n + \z)}^T : \ep \\
&= m_p (\BB_p^n)^T : \ep
\end{align*}
Note that this expression for $\ll^{P,n}$ can be taken to be the definition of total
angular momentum on particles, with $\ll_p^\BB$ being the angular momentum contribution of
particle $p$ due to $\BB_p$.

\subsubsection{Grid update}

The next step is the grid update.  Let $\GG_p = \sum_i \xt_i^{n+1} (\nabla w_{ip}^n)^T$.  Then,
\begin{align*}
\FF_p^{n+1} &= \gp{\II+\sum_i(\xt_i^{n+1} - \xx_i^n)(\nabla w_{ip}^n)^T}\FF_p^n \\
&= (\II+\GG_p-\II)\FF_p^n \\
&= \GG_p\FF_p^n
\end{align*}
For the grid update portion, we will use the following manipulations to replace cross
products with permutation tensors.
\begin{align*}
\sum_i \xx_i^n \times \AA_p \nabla w_{ip}^n
&= \sum_i \gp{\AA_p \nabla w_{ip}^n (\xx_i^n)^T}^T : \ep \\
&= \gp{\AA_p \sum_i \nabla w_{ip}^n (\xx_i^n)^T}^T : \ep \\
&= \gp{\AA_p \II}^T : \ep \\
&= \AA_p^T : \ep \\
\sum_i \xt_i^{n+1} \times \AA_p \nabla w_{ip}^n
&= \sum_i \gp{\AA_p \nabla w_{ip}^n (\xt_i^{n+1})^T}^T : \ep \\
&= \gp{\AA_p \sum_i \nabla w_{ip}^n (\xt_i^{n+1})^T}^T : \ep \\
&= \gp{\AA_p \GG_p^T}^T : \ep
\end{align*}
With this, we note the identity
\begin{align*}
\sum_i (\lambda \xt_i^{n+1} + (1-\lambda) \xx_i^n) \times \ff_i^{n+\lambda}
&= \sum_i (\lambda \xt_i^{n+1} + (1-\lambda) \xx_i^n) \times \sum_p V_p \PP_p^{n+\lambda} (\FF_p^n)^T \nabla w_{ip}^n \\
&= \sum_i (\lambda \xt_i^{n+1} + (1-\lambda) \xx_i^n) \times \sum_p V_p \FF_p^{n+\lambda} \SS_p^{n+\lambda} (\FF_p^n)^T \nabla w_{ip}^n \\
&= \sum_i \sum_p V_p \gp{\FF_p^{n+\lambda} \SS_p^{n+\lambda} (\FF_p^n)^T ((1-\lambda)\II + \lambda \GG_p)^T}^T:\ep \\
&= \sum_i \sum_p V_p \gp{\FF_p^{n+\lambda} \SS_p^{n+\lambda} ((1-\lambda)\FF_p^n + \lambda \GG_p \FF_p^n)^T}^T:\ep \\
&= \sum_i \sum_p V_p \gp{\FF_p^{n+\lambda} \SS_p^{n+\lambda} ((1-\lambda)\FF_p^n + \lambda \FF_p^{n+1})^T}^T:\ep \\
&= \sum_i \sum_p V_p \gp{\FF_p^{n+\lambda} \SS_p^{n+\lambda} (\FF_p^{n+\lambda})^T}^T:\ep \\
&= \z
\end{align*}
from which it follows that
\begin{align*}
\sum_i (\lambda \xt_i^{n+1} + (1-\lambda) \xx_i^n) \times m_i^n (\vt_i^{n+1} - \vv_i^n) &= \z.
\end{align*}
With this identity, it is finally possible to show that angular momentum is conserved
across the grid update.
\begin{align*}
\tilde{\ll}^{G,n+1} - \ll^{G,n} &= \sum_i \xt_i^{n+1} \times m_i^n \vt_i^{n+1} - \sum_i \xx_i^n \times m_i^n \vv_i^n \\
&= \sum_i \xt_i^{n+1} \times m_i^n \vt_i^{n+1} - \sum_i \xx_i^n \times m_i^n \vv_i^n - \sum_i (\lambda \xt_i^{n+1} + (1-\lambda) \xx_i^n) \times m_i^n(\vt_i^{n+1} - \vv_i^n) \\
&= \sum_i (\xt_i^{n+1} - \xx_i^n) \times m_i^n((1-\lambda) \vt_i^{n+1} + \lambda \vv_i^n) \\
&= \sum_i \dt((1-\lambda) \vt_i^{n+1} + \lambda \vv_i^n) \times m_i^n((1-\lambda) \vt_i^{n+1} + \lambda \vv_i^n) \\
&= \z
\end{align*}

\subsubsection{Transfer to particles}

Using
\begin{align*}
\ll_p^\BB &= m_p (\BB_p^{n+1})^T : \ep \\
&= m_p \gp{\frac{1}{2}\sum_i w_{ip}^n \gp{\vt_i^{n+1} (\xx_i^n-\xx_p^n+\xt_i^{n+1}-\xx_p^{n+1})^T + (\xx_i^n-\xx_p^n-\xt_i^{n+1}+\xx_p^{n+1})(\vt_i^{n+1})^T}}^T : \ep \\
&= \frac{m_p}{2} \sum_i w_{ip}^n \gp{(\xx_i^n-\xx_p^n+\xt_i^{n+1}-\xx_p^{n+1}) \times \vt_i^{n+1} + \vt_i^{n+1} \times (\xx_i^n-\xx_p^n-\xt_i^{n+1}+\xx_p^{n+1})} \\
&= \frac{m_p}{2} \sum_i w_{ip}^n \gp{(\xt_i^{n+1}-\xx_p^{n+1}) \times \vt_i^{n+1} + \vt_i^{n+1} \times (-\xt_i^{n+1}+\xx_p^{n+1})} \\
&= m_p \sum_i w_{ip}^n (\xt_i^{n+1}-\xx_p^{n+1}) \times \vt_i^{n+1} \\
&= m_p \sum_i w_{ip}^n \xt_i^{n+1} \times \vt_i^{n+1} - m_p \sum_i w_{ip}^n \xx_p^{n+1} \times \vt_i^{n+1} \\
&= m_p \sum_i w_{ip}^n \xt_i^{n+1} \times \vt_i^{n+1} - \xx_p^{n+1} \times m_p \vv_p^{n+1}
\end{align*}
we have
\begin{align*}
\ll^{P,n+1} &= \sum_p \xx_p^{n+1} \times m_p \vv_p^{n+1} + \sum_p m_p (\BB_p^{n+1})^T : \ep \\
&= \sum_p \xx_p^{n+1} \times m_p \vv_p^{n+1} + \sum_p \gp{m_p \sum_i w_{ip}^n \xt_i^{n+1} \times \vt_i^{n+1} - \xx_p^{n+1} \times m_p \vv_p^{n+1}} \\
&= \sum_p m_p \sum_i w_{ip}^n \xt_i^{n+1} \times \vt_i^{n+1} \\
&= \sum_i \xt_i^{n+1} \times \vt_i^{n+1} \sum_p w_{ip}^n m_p  \\
&= \sum_i \xt_i^{n+1} \times m_i^n \vt_i^{n+1} \\
&= \tilde{\ll}^{G,n+1}
\end{align*}
This completes the proof of angular momentum conservation.

\subsection{Stability}\label{sec:stability}

It is possible to construct a transfer that conserves angular momentum and retains affine
fields but is unstable.  This instability was observed to occur when variations in the
transfer are considered.  The instability conveniently manifests when a particle is
isolated, so the problem is easy to avoid.  We require that an isolated particle
experiencing no forces should translate uniformly with no change in $\vv_p^n$ or
$\BB_p^n$.  We now show that our scheme has this property.

Consider that there is only one particle, which experiences no forces ($\ff_i^{n+\lambda}
= \z$).  Then, the update rules for $\vv_i^n$, $\vt_i^{n+1}$, and $\xt_i^{n+1}$ reduce to
\begin{align}
\vv_i^n &= \vv_p^n + \BB_p^n (\DD_p^n)^{-1} (\xx_i^n-\xx_p^n) \\
\vt_i^{n+1} &= \vv_i^n \\
\xt_i^{n+1} &= \xx_i^n + \dt \vv_i^n
\end{align}
With these, the final particle velocity is
\begin{align}
\vv_p^{n+1} &= \sum_i w_{ip}^n \vt_i^{n+1} \\
&= \sum_i w_{ip}^n (\vv_p^n + \BB_p^n (\DD_p^n)^{-1} (\xx_i^n-\xx_p^n)) \\
&= \vv_p^n \sum_i w_{ip}^n + \BB_p^n (\DD_p^n)^{-1} \sum_i w_{ip}^n (\xx_i^n-\xx_p^n) \\
&= \vv_p^n
\end{align}
The final position is
\begin{align}
\xx_p^{n+1} &= \sum_i w_{ip}^n \xt_i^{n+1} \\
&= \sum_i w_{ip}^n (\xx_i^n + \dt \vv_i^n) \\
&= \sum_i w_{ip}^n \xx_i^n + \dt \sum_i w_{ip}^n \vv_i^n \\
&= \xx_p^n + \dt \sum_i w_{ip}^n \vt_i^{n+1} \\
&= \xx_p^n + \dt \vv_p^{n+1} \\
&= \xx_p^n + \dt \vv_p^n
\end{align}
Finally, $\BB_p^{n+1}$ is now
\begin{align}
\BB_p^{n+1} &= \frac{1}{2}\sum_i w_{ip}^n \gp{\vt_i^{n+1} (\xx_i^n-\xx_p^n+\xt_i^{n+1}-\xx_p^{n+1})^T + (\xx_i^n-\xx_p^n-\xt_i^{n+1}+\xx_p^{n+1})(\vt_i^{n+1})^T} \\
&= \frac{1}{2}\sum_i w_{ip}^n \gp{\vv_i^n (2\xx_i^n-2\xx_p^n + \dt \vv_i^n - \dt \vv_p^n)^T + (\dt \vv_p^n-\dt \vv_i^n)(\vv_i^n)^T} \\
&= \frac{1}{2}\sum_i w_{ip}^n \gp{\vv_i^n (2\xx_i^n-2\xx_p^n - \dt \vv_p^n)^T + \dt \vv_p^n(\vv_i^n)^T} \\
&= \sum_i w_{ip}^n \vv_i^n (\xx_i^n-\xx_p^n)^T - \dt \gp{\sum_i w_{ip}^n \vv_i^n} (\vv_p^n)^T + \dt \vv_p^n\gp{\sum_i w_{ip}^n \vv_i^n}^T \\
&= \sum_i w_{ip}^n \vv_i^n (\xx_i^n-\xx_p^n)^T - \dt \vv_p^n (\vv_p^n)^T + \dt \vv_p^n (\vv_p^n)^T \\
&= \sum_i w_{ip}^n \vv_i^n (\xx_i^n-\xx_p^n)^T \\
&= \sum_i w_{ip}^n (\vv_p^n + \BB_p^n (\DD_p^n)^{-1} (\xx_i^n-\xx_p^n)) (\xx_i^n-\xx_p^n)^T \\
&= \BB_p^n (\DD_p^n)^{-1} \sum_i w_{ip}^n (\xx_i^n-\xx_p^n) (\xx_i^n-\xx_p^n)^T + \vv_p^n \sum_i w_{ip}^n (\xx_i^n-\xx_p^n)^T \\
&= \BB_p^n (\DD_p^n)^{-1} \DD_p^n \\
&= \BB_p^n
\end{align}
This guarantees stability in the case of one particle.  In practice, the scheme is
observed to be stable with any number of particles when using a quadratic or cubic basis.
It is not, however, stable for a multilinear basis, as noted in
Section~\ref{sec:instability-linear}.

\subsection{Affine round trip}\label{sec:affine-round-trip}

One of the original motivations behind the original APIC scheme is that, in some
reasonable sense, it should preserve affine velocity fields.  Particles represent an
affine velocity field when $\vv_p^n = \vv + \CC \xx_p$ and $\BB_p^n = \CC \DD_p^n$ for
some vector $\vv$ and matrix $\CC$.  We require that such a velocity field be preserved in
the limit when an arbitrarily small time step is taken, so that we may assume $\dt = 0$.
The assumption $\dt = 0$ immediately implies $\vt_i^{n+1} = \vv_i^n$ and $\xt_i^{n+1} =
\xx_i^n$, from which $\xx_p^{n+1} = \xx_p^n$, $w_{ip}^{n+1} = w_{ip}^n$, and $\DD_p^{n+1}
= \DD_p^n$ follow.  The transfer to the grid simplifies to
\begin{align}
m_i^n &= \sum_p m_p w_{ip}^n \\
m_i^n \vv_i^n &= \sum_p w_{ip}^n m_p (\vv_p^n + \BB_p^n (\DD_p^n)^{-1} (\xx_i^n-\xx_p^n)) \\
&= \sum_p w_{ip}^n m_p (\vv + \CC \xx_p + \CC \DD_p^n (\DD_p^n)^{-1} (\xx_i^n-\xx_p^n)) \\
&= \sum_p w_{ip}^n m_p (\vv + \CC \xx_i^n) \\
&= m_i^n (\vv + \CC \xx_i^n) \\
\vv_i^n &= \vv + \CC \xx_i^n
\end{align}
so that the grid velocity field is produced by the same affine velocity field.
\begin{align}
\vv_p^{n+1} &= \sum_i w_{ip}^n \vt_i^{n+1} \\
&= \sum_i w_{ip}^n \vv_i^n \\
&= \sum_i w_{ip}^n (\vv + \CC \xx_i^n) \\
&= \vv \sum_i w_{ip}^n + \CC \sum_i w_{ip}^n \xx_i^n \\
&= \vv + \CC \xx_p^n \\
&= \vv + \CC \xx_p^{n+1} \\
\BB_p^{n+1} &= \frac{1}{2}\sum_i w_{ip}^n \gp{\vt_i^{n+1} (\xx_i^n-\xx_p^n+\xt_i^{n+1}-\xx_p^{n+1})^T + (\xx_i^n-\xx_p^n-\xt_i^{n+1}+\xx_p^{n+1})(\vt_i^{n+1})^T} \\
&= \sum_i w_{ip}^n \vv_i^n (\xx_i^n-\xx_p^n)^T \\
&= \sum_i w_{ip}^n (\vv + \CC \xx_i^n) (\xx_i^n-\xx_p^n)^T \\
&= \vv \sum_i w_{ip}^n (\xx_i^n-\xx_p^n)^T + \CC \sum_i w_{ip}^n (\xx_i^n-\xx_p^n) (\xx_i^n-\xx_p^n)^T + \CC \xx_p^n \sum_i w_{ip}^n (\xx_i^n-\xx_p^n)^T \\
&= \CC \DD_p^n \\
&= \CC \DD_p^{n+1}
\end{align}
The new particle state corresponds to the same affine velocity field, so the field has
been preserved across the transfers.

\subsection{Unifying PIC, RPIC and APIC}\label{sec:unified}

For each of PIC, RPIC and APIC the transfer from particle to grid can be written as
$m_{ip}=m_pN(\xx_p-\xx_i)$, $m_i=\sum_p m_{ip}$,
$(m\vv)_{ip}=m_{ip}(\vv_p+\CC_p(\xx_i-\xx_p))$ and $(m\vv)_i=\sum_p(m\vv)_{ip}$, where the
$\CC_p$ is zero, skew or a full matrix to distinguish PIC, RPIC and APIC
respectively. However, when designing the transfer back from grid to particle, the details
are less obviously related. There is, in fact, a description that unifies RPIC, APIC and
PIC. It starts with the alternative notation
$$
\vv_{ip}=\sum_{j=1}^{N_r} s_{jp} \bb_{jpi}
$$
to describe the velocity field local to the particle. Here, the
$\bb_{jpi}\in\mathbb{R}^3$, $\bb_{jp}\in\mathbb{R}^{3{N_g}}$ with $N_g$ equal to the
number of grid nodes and
$$
\bb_{jp}=\left(\begin{array}{c}
\bb_{jp1}\\
\bb_{jp2}\\
\vdots\\
\bb_{jpN_G}
\end{array}
\right).
$$
The $\bb_{jp}$ form a reduced basis for the grid velocity field $\vv_{ip}$ local to
particle $p$. That is, the $\bb_{jp}\in\mathbb{R}^{3{N_g}}$ are individual modes defined
over the grid and the $s_{jp}$ describe the local particle state, e.g. they are equivalent
to $\vv_p$ and $\CC_P$ for APIC. The choice of the basis vectors
$\bb_{jp}\in\mathbb{R}^{3{N_g}}$ is what distinguishes PIC from RPIC from APIC etc. For
example, PIC uses $N_r=3$ and
$$
\bb_{jp}=\left(\begin{array}{c}
\ee_j\\
\ee_j\\
\vdots\\
\ee_j
\end{array}
\right) \ \ \textrm{and} \ \
\vv_{p}=\left(\begin{array}{c}
s_{1p}\\
s_{2p}\\
s_{3p}
\end{array}
\right)
$$
for $j=1,2,3=N_r$ with $\ee_j\in\mathbb{R}^3$ the $j^\textrm{th}$ standard basis vector
for $\mathbb{R}^3$. RPIC uses $N_r=6$ with the same $\bb_{jp}$ as PIC for $j=1,2,3$ and
$$
\bb_{jp}=\left(\begin{array}{c}
\sum_{k=1}^3\epsilon_{j-3k1}r_{1pk}\\
\sum_{k=1}^3\epsilon_{j-3k2}r_{1pk}\\
\sum_{k=1}^3\epsilon_{j-3k3}r_{1pk}\\
\sum_{k=1}^3\epsilon_{j-3k1}r_{2pk}\\
\vdots\\
\sum_{k=1}^3\epsilon_{j-3k1}r_{N_gpk}\\
\sum_{k=1}^3\epsilon_{j-3k2}r_{N_gpk}\\
\sum_{k=1}^3\epsilon_{j-3k3}r_{N_gpk}\\
\end{array}
\right)
$$
for $j=4,5,6=N_r$ where $\epsilon_{ijk}$ is the permutation tensor (such that the
$k^\textrm{th}$ component of $\mb{a}\times\mb{b}$ is $\sum_{i,j}\epsilon_{ijk}a_ib_j$) and
$r_{ipk}$ is the $k^\textrm{th}$ component of $\rr_{ip}=\xx_i-\xx_p$. With this
convention,
$$
\vv_{p}=\left(\begin{array}{c}
s_{1p}\\
s_{2p}\\
s_{3p}
\end{array}
\right)
 \ \ \textrm{and} \ \
\boldsymbol\omega_{p}=\left(\begin{array}{c}
s_{4p}\\
s_{5p}\\
s_{6p}
\end{array}
\right).
$$
Lastly, APIC uses the same $\bb_{jp}$ as RPIC for $j=1,2,\hdots,6$ and
$$
\bb_{jp}=\left(\begin{array}{c}
\sum_{k=1}^3|\epsilon_{j-6k1}|r_{1pk}\\
\sum_{k=1}^3|\epsilon_{j-6k2}|r_{1pk}\\
\sum_{k=1}^3|\epsilon_{j-6k3}|r_{1pk}\\
\sum_{k=1}^3|\epsilon_{j-6k1}|r_{2pk}\\
\vdots\\
\sum_{k=1}^3|\epsilon_{j-6k1}|r_{N_gpk}\\
\sum_{k=1}^3|\epsilon_{j-6k2}|r_{N_gpk}\\
\sum_{k=1}^3|\epsilon_{j-6k3}|r_{N_gpk}\\
\end{array}
\right)
$$
for $j=7,8,9$ (which represent symmetric matrices with zero diagonal) and
$$
\bb_{jp}=\left(\begin{array}{c}
\ee_{j-9}\ee_{j-9}^T\rr_{1p}\\
\ee_{j-9}\ee_{j-9}^T\rr_{2p}\\
\vdots\\
\ee_{j-9}\ee_{j-9}^T\rr_{N_Gp}\\
\end{array}
\right)
$$
for $j=10,11,12=N_r$ which represent the diagonal matrices. With this convection,
$$
\vv_{p}=\left(\begin{array}{c}
s_{1p}\\
s_{2p}\\
s_{3p}
\end{array}
\right)
 \ \ \textrm{and} \ \
\CC_{p}=\left(\begin{array}{ccc}
s_{10p}&s_{7p}-s_{4p}&s_{8p}-s_{5p}\\
s_{4p}+s_{7p}&s_{11p}&s_{9p}-s_{6p}\\
s_{5p}+s_{8p}&s_{6p}+s_{9p}&s_{12p}
\end{array}
\right).
$$
\subsubsection{Transfer from particle to grid}
With this notation, the transfer from particle to grid is $m_{ip}=m_pN(\xx_p-\xx_i)$,
$m_i=\sum_p m_{ip}$, $(m\vv)_{ip}=m_{ip}\sum_{j=1}^{N_r} s_{jp} \bb_{jpi}$ and
$(m\vv)_i=\sum_p(m\vv)_{ip}$. That is, the grid momenta are just the sum of the momenta
modes local to each particle. Notably, this describes the PIC, RPIC and APIC transfers in
one description. If we define the total linear momentum of the particle state to be the
sum of the total linear momenta from each local particle state, and the total angular
momentum of particle state to be the sum of the total angular momenta from each local
particle state (computed about the particle) plus the angular momenta of the particles,
then the transfer conserves linear and angular momentum by reasoning analogous to that in
Section \ref{sec:RPICproofs}.

\subsubsection{Transfer from grid to particle}
Using this notation, the transfer from grid to particle is done by determining
$\tilde{s}_{jp}$ from the updated grid velocities $\tilde{\vv}_i$. We can do this in a way
that conserves linear, angular momenta, as well as generalized moments directly by solving
the system
$$ \bb_{jp}^T\left(\begin{array}{c}
  m_{1p}\tilde{\vv}_1\\ m_{2p}\tilde{\vv}_2\\ \vdots\\ m_{N_gp}\tilde{\vv}_{N_g}
\end{array}\right)=\bb_{jp}^T\MM\bb_{ip}\tilde{s}_{ip}
$$
for $\tilde{s}_{ip}$. Notably, this describes the PIC and RPIC transfers when the $N_r=3$
and $N_r=6$ respectively. Furthermore, it generalizes the result to the affine
case. Remarkably, it can be shown that the matrix $\bb_{jp}^T\MM\bb_{ip}$ is both diagonal
and constant in time for quadratic and cubic B-splines, i.e. does not depend on the
configuration of the particles relative to the grid. This not only means that these solves
can be done efficiently, but it also shows that for APIC, $\tilde{s}_{ip}$ are the PIC
modes $i=1,2,3$ and the RPIC modes for $i=4,5,6$ and that the remaining APIC modes are
determined independently since the components are decoupled in the solve. Lastly, the
transfer conserves linear and angular momentum by the same argument as for RPIC since the
right hand side terms
$$
\bb_{jp}^T\left(\begin{array}{c}
m_{1p}\tilde{\vv}_1\\
m_{2p}\tilde{\vv}_2\\
\vdots\\
m_{N_gp}\tilde{\vv}_{N_g}
\end{array}\right)=\bb_{jp}^T\MM\bb_{ip}\tilde{s}_{ip}
$$
are the linear momentum components for $j=1,2,3$ and the angular momentum components
(computed about the particle) for $j=4,5,6$.

\begin{figure}
\begin{center}
\includegraphics[draft=\mydraft]{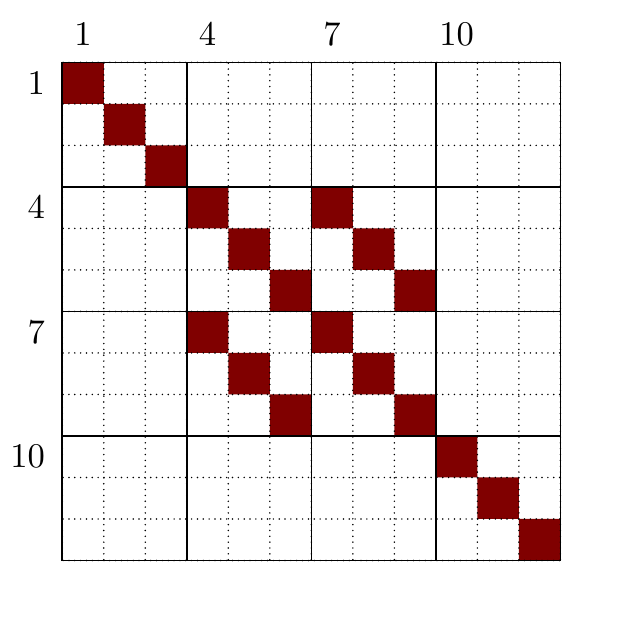}\vspace{-1.5em}
\end{center}
\caption{The matrix structure of $\bb_{jp}^T\MM\bb_{ip}$ in the multilinear case.\label{fig:multilinearstructure}}
\end{figure}

\subsubsection{Coefficient computations}
To construct the transfer, we need to compute the basis coefficients $\tilde{s}_{ip}$.
This requires building the matrix $\bb_{jp}^T\MM\bb_{ip}$ and inverting it. Notably, building this
matrix and its inverse requires very little computation. In the case of cubic and quadratic
B-spline interpolation, the matrix is actually constant and diagonal. For cubic B-splines,
$\bb_{jp}^T\MM\bb_{ip}$ is diagonal with entries \begin{equation*} m_p
\times\{1,1,1,\frac{2}{3}\dx^2,\frac{2}{3}\dx^2,\frac{2}{3}\dx^2,\frac{2}{3}\dx^2,\frac{2}{3}\dx^2,\frac{2}{3}\dx^2,\frac{1}{3}\dx^2,\frac{1}{3}\dx^2,\frac{1}{3}\dx^2\} \end{equation*}
For quadratic B-splines, $\bb_{jp}^T\MM\bb_{ip}$ is diagonal with entries \begin{equation*} m_p
\times\{1,1,1,\frac{1}{2}\dx^2,\frac{1}{2}\dx^2,\frac{1}{2}\dx^2,\frac{1}{2}\dx^2,\frac{1}{2}\dx^2,\frac{1}{2}\dx^2,\frac{1}{4}\dx^2,\frac{1}{4}\dx^2,\frac{1}{4}\dx^2\}. \end{equation*}
For the cubic and quadratic cases, this transfer is equivalent to those derived in
previous sections, albeit without the intuition needed to prove a number of the useful
properties.

For multilinear interpolation function, it is a symmetric matrix (but not diagonal).
\begin{align*}
\bb_{jp}^T\MM\bb_{ip} &= m_p \mx{
  \II & \z & \z & \z \\
  \z & \AA_0 & \AA_1 & \z \\
  \z & \AA_1 & \AA_0 & \z \\
  \z & \z & \z & -\dx \ZZ - \ZZ^2
}
\end{align*}
where
\begin{align*}
\AA_0 = -\mbox{diag}\mx{z_{p2}^2+z_{p3}^2+{\dx} (z_{p2}+z_{p3}) \\ z_{p1}^2+z_{p3}^2+{\dx} (z_{p1}+z_{p3}) \\ z_{p1}^2+z_{p2}^2+{\dx} (z_{p1}+z_{p2})}
\qquad
\AA_1 = -\mbox{diag}\mx{({\dx}+z_{p2}+z_{p3}) (z_{p2}-z_{p3}) \\ ({\dx}+z_{p1}+z_{p3}) (z_{p3}-z_{p1}) \\ ({\dx}+z_{p1}+z_{p2}) (z_{p1}-z_{p2})},
\end{align*}
$\OO_p$ is the the bottom left corner location of the cell that particle
$p$ affects, $\zz_p = \OO_p - \xx_p$ and $\ZZ =
\mbox{diag}(\zz)$. Figure~\ref{fig:multilinearstructure} shows the structure of
this matrix. Its inverse has the same structure.

\subsection{Degrades to backward Euler case}

The transfers are the same as in \cite{jiang:2015:apic}, except for the update rule for
$\BB_p^{n+1}$, which we show below simplifies into the transfer from
\cite{jiang:2015:apic} with $\lambda=0$.  Using $\lambda$ recovers the the backward Euler
grid update rule $\xt_i^{n+1} = \xx_i^n + \dt\vt_i^{n+1}$.  Then,
\begin{align*}
\xt_i^{n+1} &= \xx_i^n + \dt\vt_i^{n+1} \\
\sum_i w_{ip}^n \xt_i^{n+1} &= \sum_i w_{ip}^n (\xx_i^n + \dt\vt_i^{n+1}) \\
\xx_p^{n+1} &= \xx_p^n + \dt \vv_p^{n+1} \\
\BB_p^{n+1} &= \frac{1}{2}\sum_i w_{ip}^n \gp{\vt_i^{n+1} (\xx_i^n-\xx_p^n+\xt_i^{n+1}-\xx_p^{n+1})^T + (\xx_i^n-\xx_p^n-\xt_i^{n+1}+\xx_p^{n+1})(\vt_i^{n+1})^T} \\
&= \frac{1}{2}\sum_i w_{ip}^n \gp{\vt_i^{n+1} (2\xx_i^n - 2\xx_p^n + \dt\vt_i^{n+1} - \dt \vv_p^{n+1})^T + \dt (\vv_p^{n+1}-\vt_i^{n+1})(\vt_i^{n+1})^T} \\
&= \sum_i w_{ip}^n \vt_i^{n+1} (\xx_i^n - \xx_p^n)^T + \frac{\dt}{2}\sum_i w_{ip}^n \gp{-\vt_i^{n+1} (\vv_p^{n+1})^T + \vv_p^{n+1}(\vt_i^{n+1})^T} \\
&= \sum_i w_{ip}^n \vt_i^{n+1} (\xx_i^n - \xx_p^n)^T + \frac{\dt}{2} \gp{-\vv_p^{n+1} (\vv_p^{n+1})^T + \vv_p^{n+1}(\vv_p^{n+1})^T} \\
&= \sum_i w_{ip}^n \vt_i^{n+1} (\xx_i^n - \xx_p^n)^T
\end{align*}
This was the original APIC transfer.

Another departure from \cite{jiang:2015:apic} is the update rule
for $\xx_p^{n+1}$.  We note, however, that these are also equivalent in the $\lambda = 0$
case.
\begin{align*}
\xx_p^{n+1} &= \sum_i w_{ip}^n \xt_i^{n+1}  \\
&= \sum_i w_{ip}^n (\xx_i^n + \dt \lambda \vv_i^n + \dt (1-\lambda)  \vt_i^{n+1}) \\
&= \xx_p^n+ \dt \lambda \sum_i w_{ip}^n\vv_i^n + \dt (1-\lambda)\sum_i w_{ip}^n\vt_i^{n+1} \\
&= \xx_p^n+ \dt \lambda \sum_i w_{ip}^n\vv_i^n + \dt (1-\lambda)\vv_p^{n+1}
\end{align*}
Note that the second term is not $\vv_p^n$.  This term vanishes in the special case
$\lambda=0$, so that the transfer in \cite{jiang:2015:apic} could be done using the
particle velocity.  We see that the proposed method represents a generalization of the
original APIC scheme.

\subsection{Note on transfer construction}\label{sec:APIC_derivation}

The most important difference between the method described above and the original APIC
method from \cite{jiang:2015:apic} is \eqref{eqn:update-bb}.  This transfer was
constructed by first assuming that the transfer should take the form
\begin{align*}
\BB_p^{n+1} &= \sum_i w_{ip}^n \gp{\vt_i^{n+1} (a \xx_i^n+b\xx_p^n+c\xt_i^{n+1}+d\xx_p^{n+1})^T + (e\xx_i^n+f\xx_p^n+g\xt_i^{n+1}+h\xx_p^{n+1})(\vt_i^{n+1})^T}.
\end{align*}
Terms involving $\vv_i^n$ could also be considered; such terms would add a FLIP-like
character to the transfer.  Terms similar to $(\vt_i^{n+1})^T \xx_i^n \II$ are also
technically possible; we do not include them since we did not find them to contribute
meaningfully to the transfer.  We restrict ourselves here to the form above, which leaves
us to choose the eight coefficients.  We require (1) angular momentum conservation (See
Section~\ref{sec:angular-momentum}), (2) affine round trip (See
Section~\ref{sec:affine-round-trip}), and (3) one particle stability (See
Section~\ref{sec:stability}).  These three constraints uniquely determine all eight
coefficients in the general case.  In the special case $\lambda = 0$, the eight terms are
not linearly independent; this allows additional freedom to eliminate terms, resulting in
the simpler transfer from \cite{jiang:2015:apic}.

\begin{figure}
\begin{center}
\includegraphics[draft=\mydraft,width=.49\textwidth]{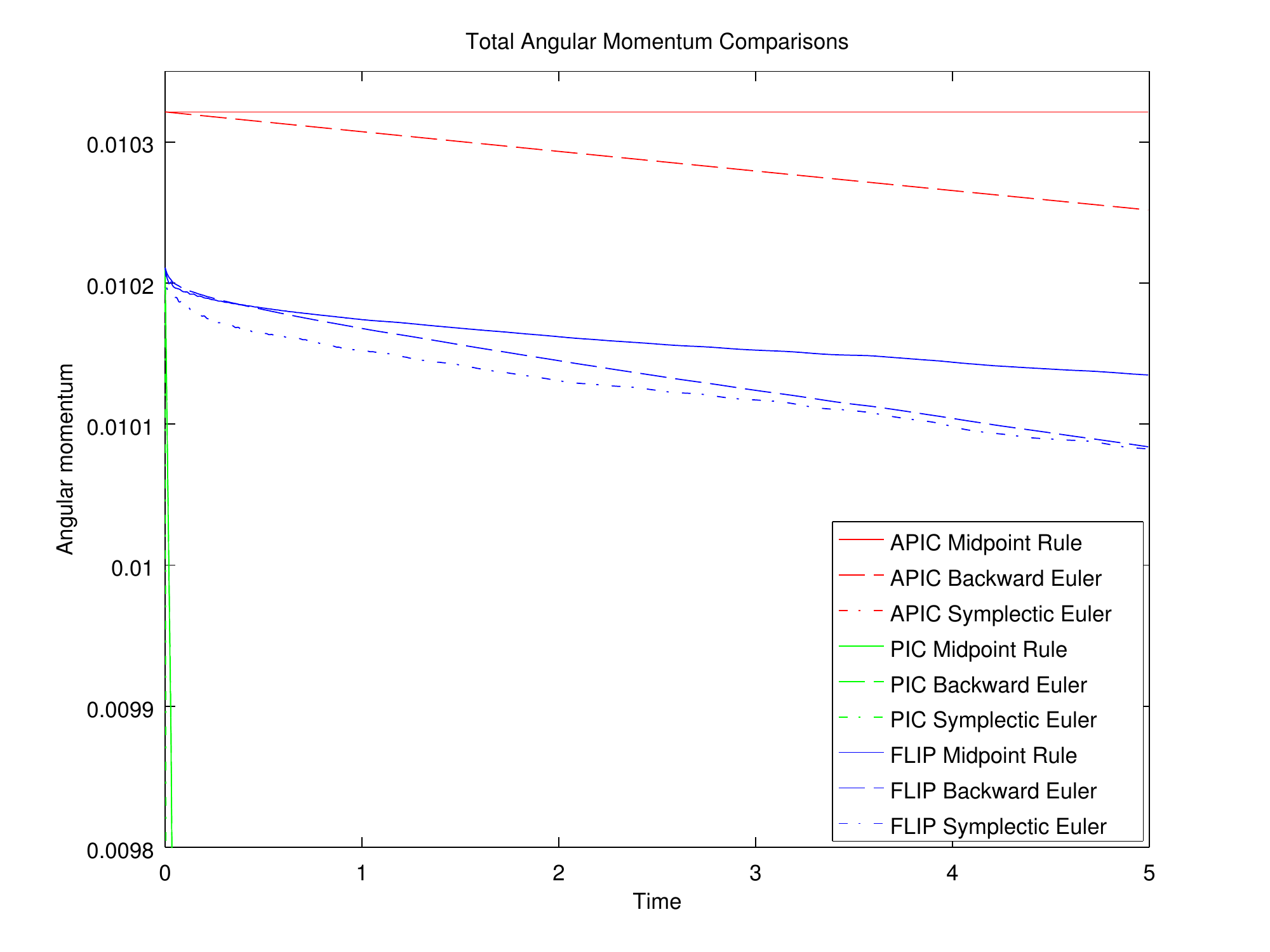}
\includegraphics[draft=\mydraft,width=.49\textwidth]{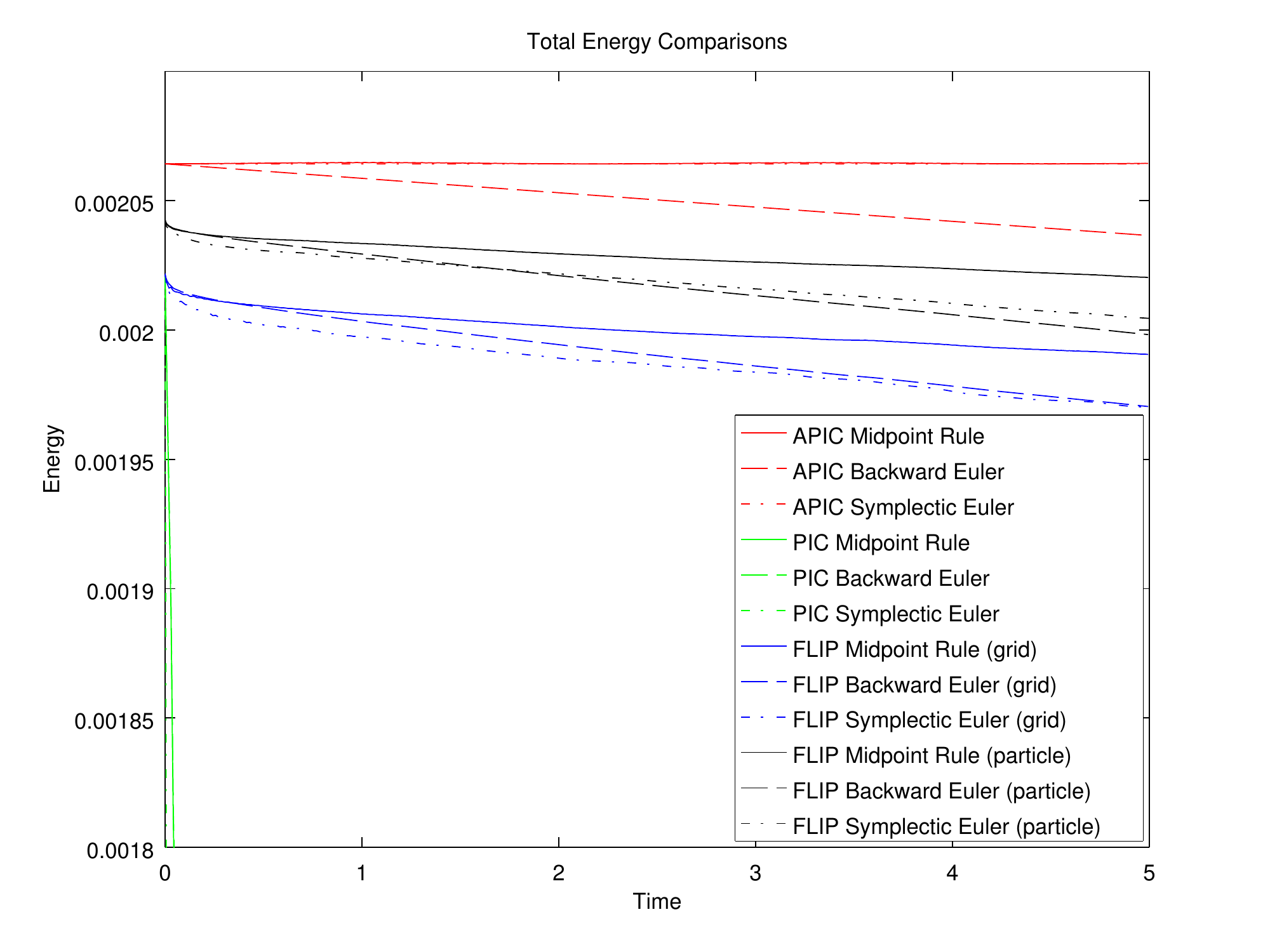}
\end{center}
\caption{Rotation test.\label{fig:rotation}}
\end{figure}

\subsection{Stability concerns for multilinear interpolation}\label{sec:instability-linear}

\begin{figure}[bt]
\begin{center}
\includegraphics[draft=\mydraft,width=.49\textwidth]{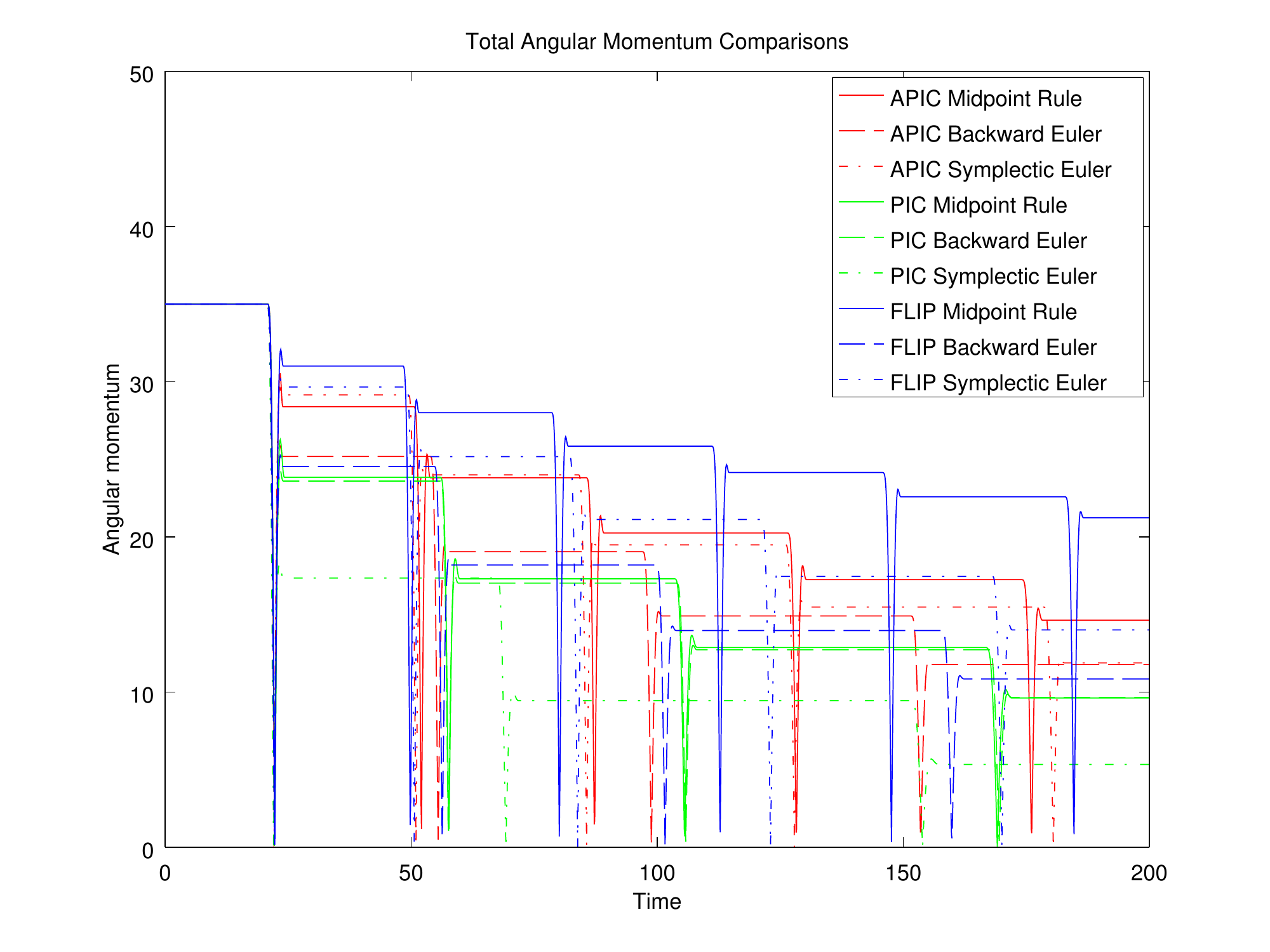}
\includegraphics[draft=\mydraft,width=.49\textwidth]{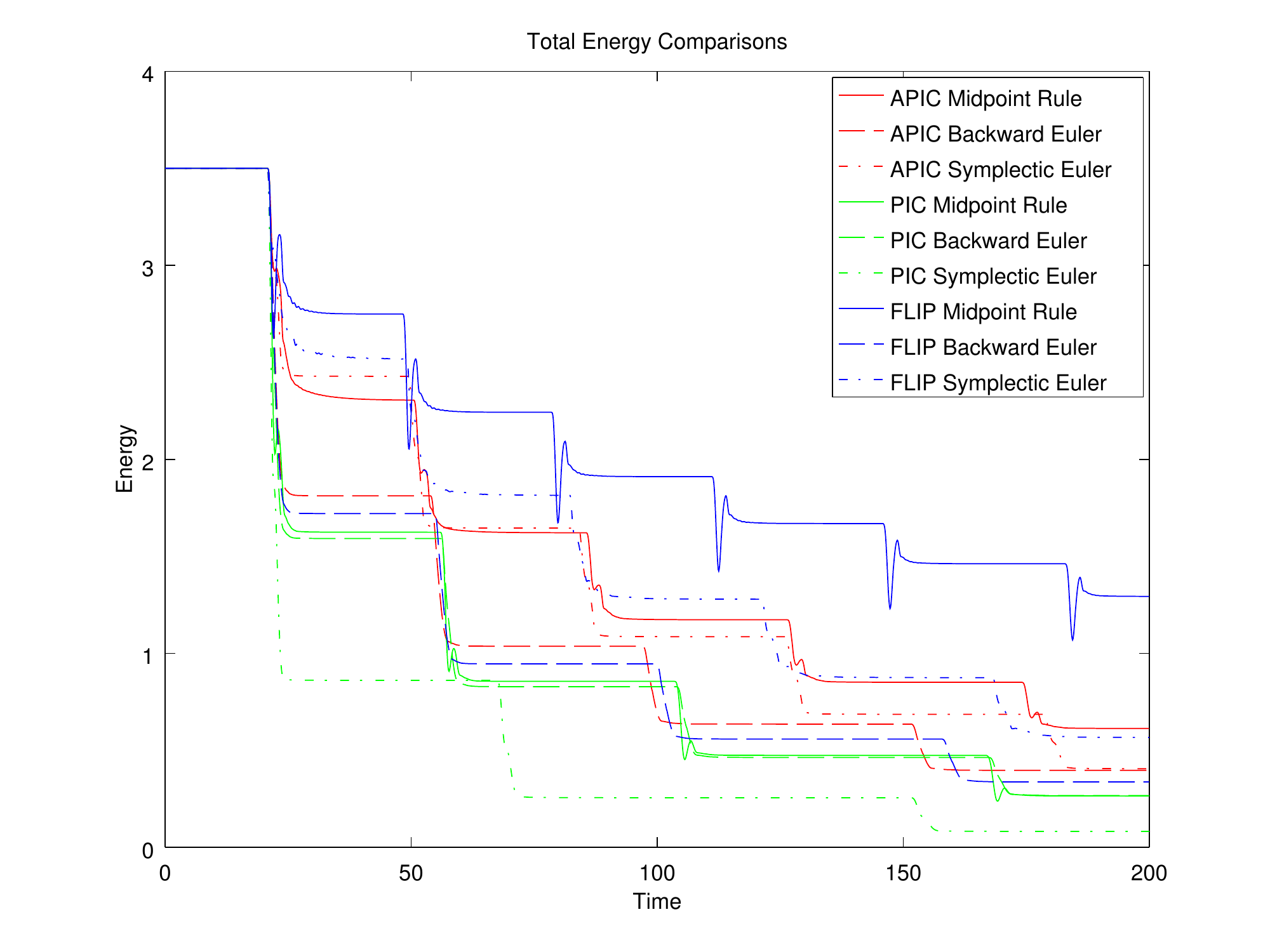}
\end{center}
\caption{Rebound test.\label{fig:rebound}}
\end{figure}

With linear interpolation weights, $\DD_p^n$ is not invertible. The particle to grid transfer from
\begin{align*}
m_i^n \vv_i^n &= \sum_p w_{ip}^n m_p (\vv_p^n + \BB_p^n (\DD_p^n)^{-1} (\xx_i^n-\xx_p^n))
\end{align*}
can in the multilinear interpolation case be re-written as
\begin{align*}
m_i^n \vv_i^n &= \sum_p (w_{ip}^n m_p \vv_p^n + m_p \BB_p^n \nabla w^n_{ip})
\end{align*}
using $w^n_{ip} (\DD_p^n)^{-1} (\xx_i^n-\xx_p^n) = \nabla w^n_{ip}$.  To see why this can
cause problems, consider the case with one particle. Let $\vv_p^n=\z$ and
$\BB_p^n=\II$. Then
\begin{align*}
m_i^n \vv_i^n &= m_p \nabla w^n_{ip} \\
\vv_i^n &= \frac{\nabla w^n_{ip}}{w^n_{ip}}
\end{align*}
Consider a grid cell at $[0,\dx] \times [0,\dx]$ with grid degrees of freedom at
$\xx_{(i,j)} = (i \dx, j \dx)$.  If the particle $p$ is at $(\epsilon \dx,\frac{1}{2}
\dx)$, then $w_{(1,1)p}^n = \frac{\epsilon}{2}$ and $\nabla w_{(1,1)p}^n = \langle
\frac{1}{2\dx} , \frac{\epsilon}{\dx} \rangle$.  But then, $\vv_{(1,1)}^n = \langle
\frac{1}{\epsilon\dx} , \frac{2}{\dx} \rangle$, which is unbounded.  This in turn results
in a kinetic energy contribution of $\frac{1}{2} m_{(1,1)}^n \|\vv_{(1,1)}^n\|^2 =
\frac{1}{4 \dx} m_p (\epsilon^{-1} + 4 \epsilon)$.  Since $\epsilon$ can be arbitrarily
small, the energy of the grid node can be arbitrarily large.  This unbounded growth in
energy causes instability and makes a multilinear interpolation kernel unsuitable for this
APIC formulation.

\begin{figure}
\begin{center}
\includegraphics[draft=\mydraft,width=.49\textwidth]{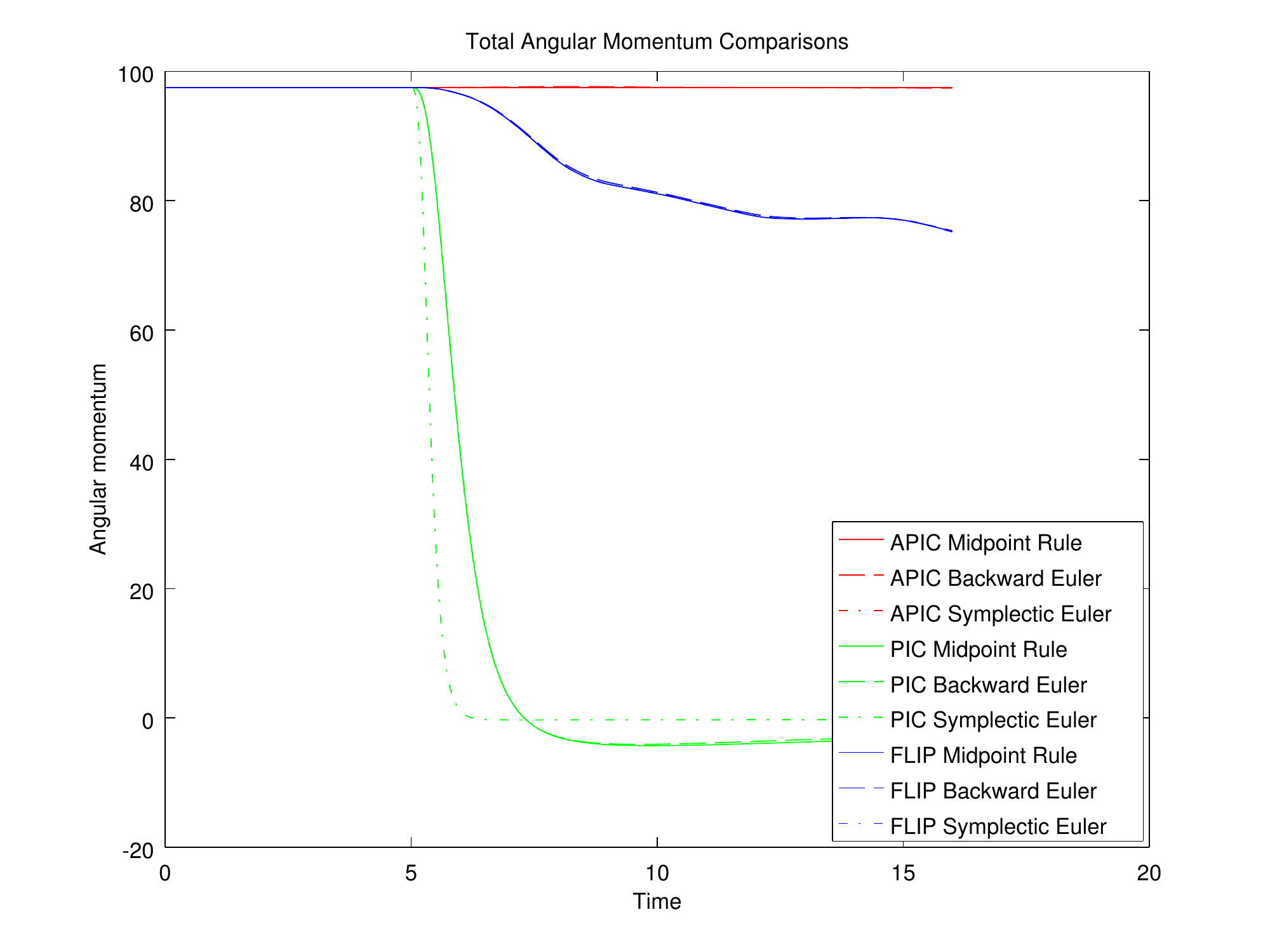}
\includegraphics[draft=\mydraft,width=.49\textwidth]{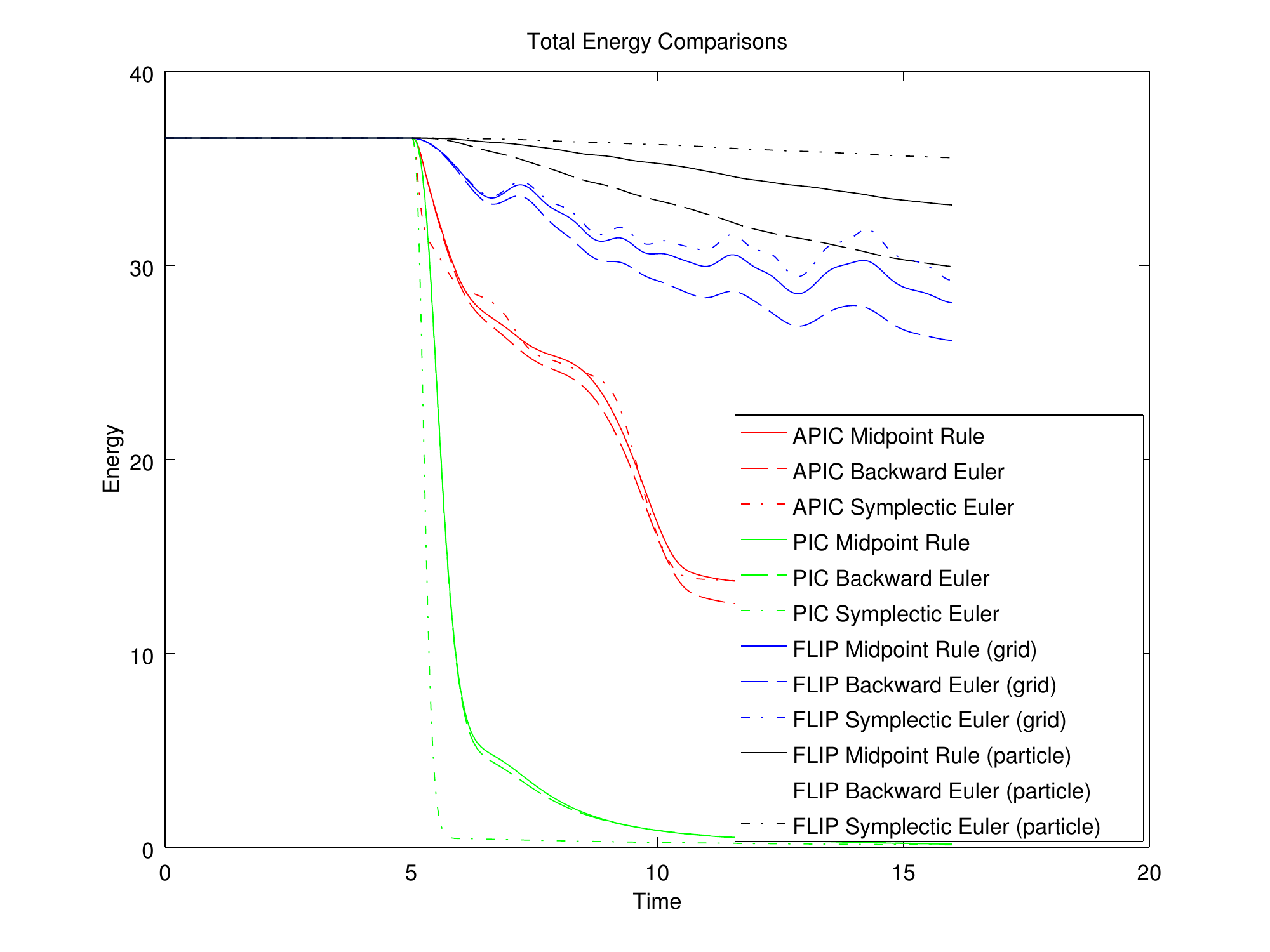}
\end{center}
\caption{Skew impact of two elastic cylinders.\label{fig:skew}}
\end{figure}

In order to use APIC using multilinear interpolation function without being unstable,
we can lag the affine matrix with the transfers being
\begin{align*}
    m_i^n \vv_i^n &= \sum_p w_{ip}^n m_p (\vv_p^n + \CC_p^n  (\xx_i^n-\xx_p^n)), \\
    \CC_p^{n+1} &= \sum_i w_{ip}^n \vt_i ((\DD_p^n)^{-1} (\xx_i^n-\xx_p^n))^T.
\end{align*}
For multilinear interpolation, it further simplifies to $\CC_p^{n+1}= \sum_i \vt_i (\nabla
w_{ip}^n)^T$.  Note that this formulation does not suffer from the same energy increasing
problem as long as $\CC$ is bounded.  The difference is effectively that the $\BB$
formulation inverts $\DD_p^n$ at the end of the time step rather than doing so at the
beginning of the next time step.  In the quadratic and cubic cases, the $\CC$ formulation
and the $\BB$ formulation are equivalent, since $\DD_p^n$ is a constant scalar multiple of
the identity and thus $\DD_p^n = \DD_p^{n+1}$. For multilinear interplation, we always use
the lagged version for stability.

%

\section{Numerical simulations}

\subsection{Rotating elastic cylinder}
We begin our tests by running a simple
rotation test.  We use a $[0,1] \times [0,1]$ domain with $32 \times 32$
resolution.  We initialize a circle with radius $0.3$ centered at $(0.5,0.5)$,
seeded with four particles per cell.  The circle begins rotating with angular
velocity $0.4$ about its center.  We use an initial density $\rho = 2$ and a
Neo-Hookean constitutive model with $E = 1000$ and $\nu = 0.3$. See
Figure~\ref{fig:rotation}.

\subsection{Rebound of an elastic cylinder}
We run the same example as in section 4.1 of \cite{love:2006:stable-mpm}.

The grid spacing is $h=0.5$. Slip boundary conditions are applied at $x=0$ and $x=15$. The
cylinder is initially centered at $(2.5,2.5)$ and has radius $1.5$. MPM particles are
sampled with alignment to the grid with spacing $0.25$ (so $4$ particles per cell for a full
cell). Material density is $4$.
The constitutive model is Neo-Hookean with Young's Modulus
$85.5$ and Poisson's ratio $0.425$. The initial velocity of the cylinder is $(0.5,0)$.
See Figure~\ref{fig:rebound}.

\subsection{Skew impact of two elastic cylinders}

We run the same example as in section 4.2 of \cite{love:2006:stable-mpm}.

The grid spacing is $h=1$. The first cylinder is initially centered at $(3,3)$ with
velocity $(0.75,0)$. The second cylinder is initially centered at $(16,5)$ with
velocity $(-0.75,0)$. Each cylinder has radius 2. MPM particles are sampled with
alignment to the grid with spacing $0.5$ (so $4$ particles per cell for a full cell).
Material density is $5$.  The constitutive model is Neo-Hookean with Young's Modulus
$31.685$ and Poisson's ratio $0.44022$.

\begin{figure}
\begin{center}
\includegraphics[draft=\mydraft,width=.49\textwidth]{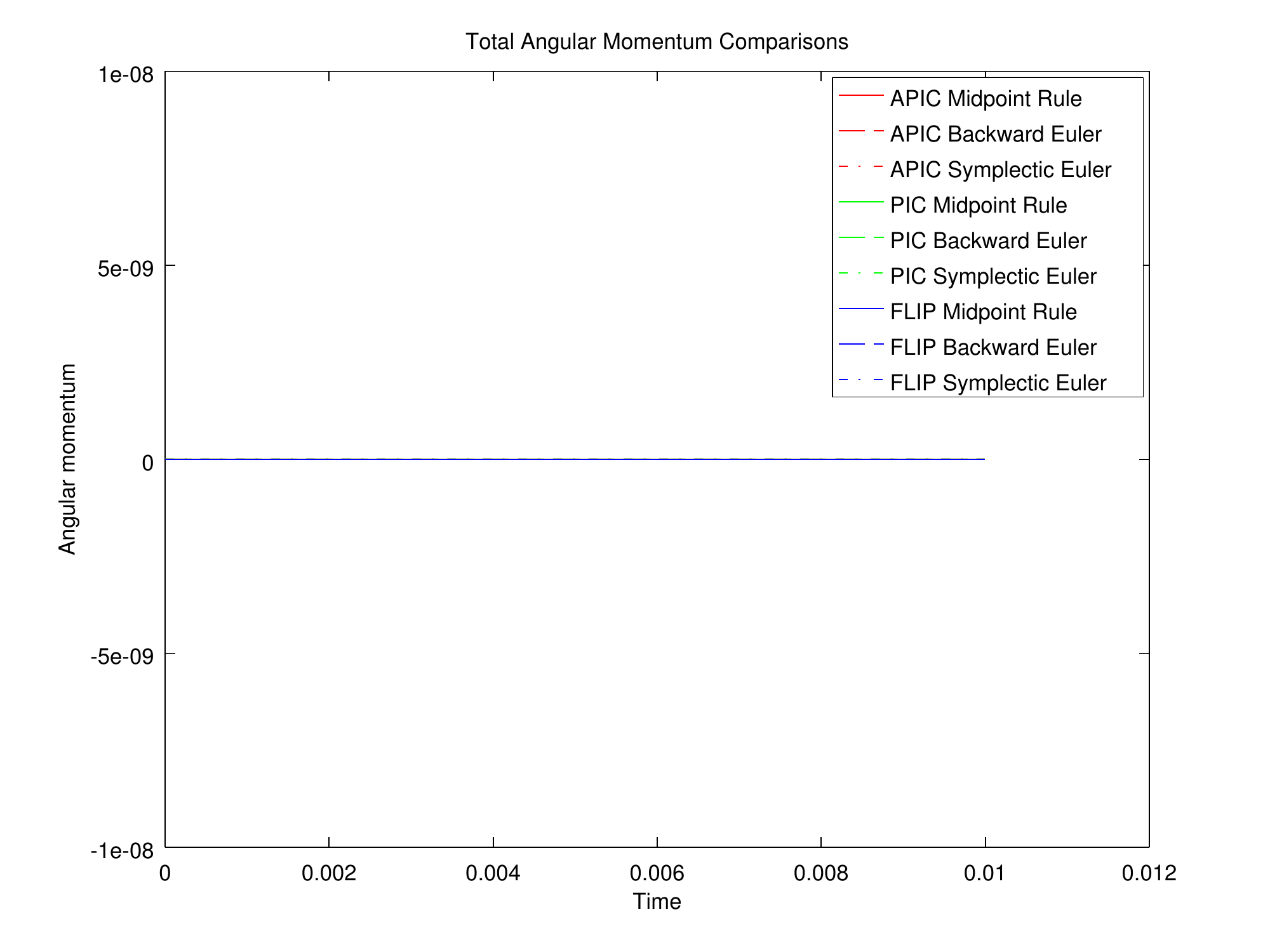}
\includegraphics[draft=\mydraft,width=.49\textwidth]{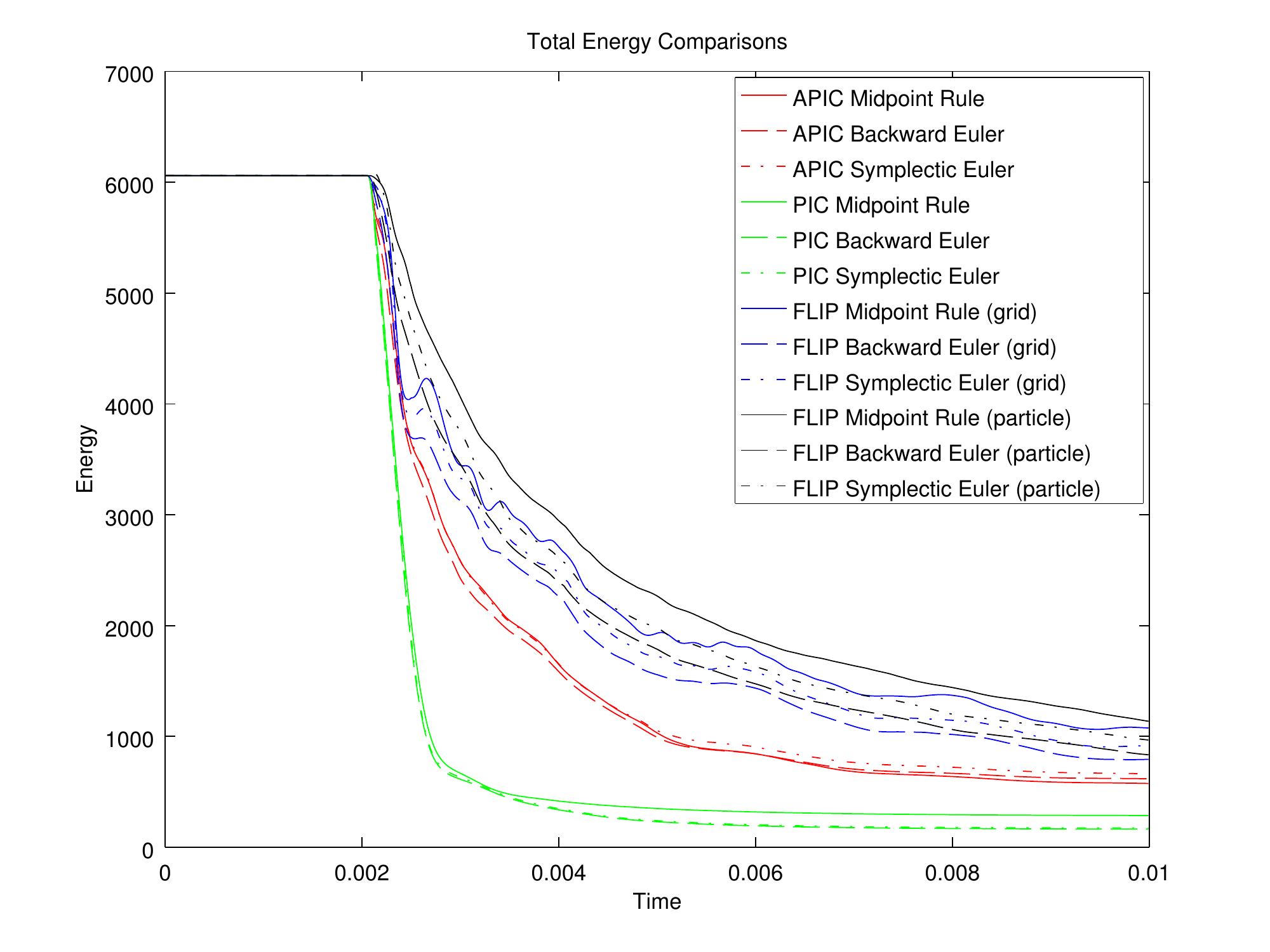}
\end{center}
\caption{Colliding rings.\label{fig:rings}}
\end{figure}

\begin{figure}
\begin{center}
\includegraphics[draft=\mydraft,width=\textwidth]{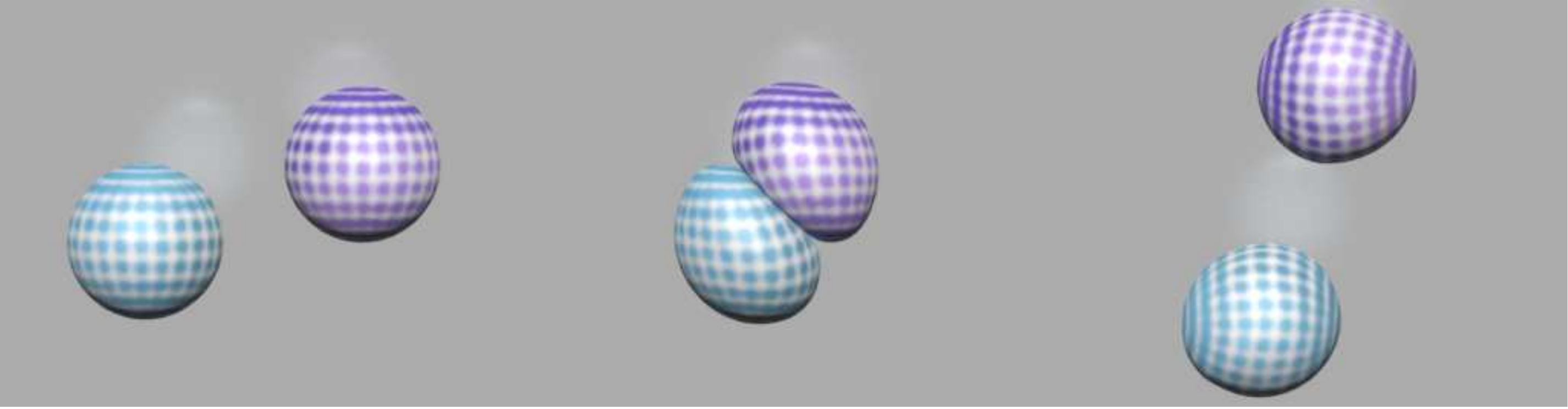}
\caption{Colliding spheres 3D, frame $70/145/253$ with framerate $24 Hz$.\label{fig:spheres-render}}
\end{center}
\end{figure}

\subsection{Elastic cylinder collision}
We extend the previous example to two colliding hollow cylinders.

The grid spacing is $h=0.01$. The first ring is initially centered at $(0.1,0.24)$ with
velocity $(50,0)$. The second ring is initially centered at $(0.4,0.24)$ with velocity
$(-50,0)$. Each ring has outer radius $0.04$ and inner radius $0.03$.
MPM particles are sampled with alignment to the grid with spacing $1/300$. Material density
is $1010$.  The constitutive model is Neo-Hookean with Young's Modulus $7.3$e$7$ and Poisson's
ratio $0.4$.

\subsection{Elastic sphere collision (3D)}
We extend the skew impact of spheres to 3D.
The grid spacing is $h=30/256$. The first sphere is initially centered at $(10,13,15)$
with velocity $(0.75,0,0)$. The second sphere is initially centered at $(20,15,15)$
with velocity $(-0.75,0,0)$. Each sphere has radius $2$. MPM particles are sampled with 4
particles per cell for  a total particle count of $333,213$.  Material density is $5$.
The constitutive model is Neo-Hookean with Young's Modulus $31.685$ and Poisson's ratio
$0.44022$.  Figure \ref{fig:spheres-render} shows the visualized objects at time $2.92$, $6.04$
and $10.54$.

\begin{figure}
\begin{center}
\includegraphics[draft=\mydraft,width=\textwidth]{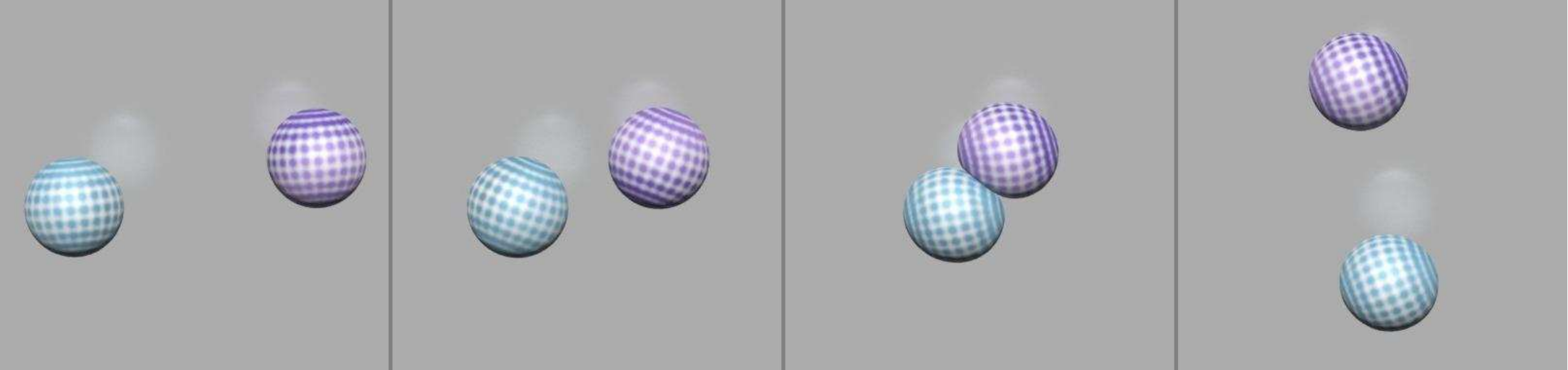}
\caption{Colliding spheres 3D with initial angular velocity of $1$, frame $1/34/63/92$ with framerate $12 Hz$.\label{fig:spheres_angular-render}}
\end{center}
\end{figure}

We further extend the previous test by initializing each sphere with an angular velocity of $(0,0,1)$
(i.e., the spheres initially rotate counterclockwise) and scaling the Young's modulus by $8$.
Figure \ref{fig:spheres_angular-render} shows the visualized objects at time
$0.08$, $2.83$, $5.25$ and $7.67$.

\subsection{Torus dropping}

\begin{figure}
\begin{center}
\includegraphics[draft=\mydraft,width=\textwidth]{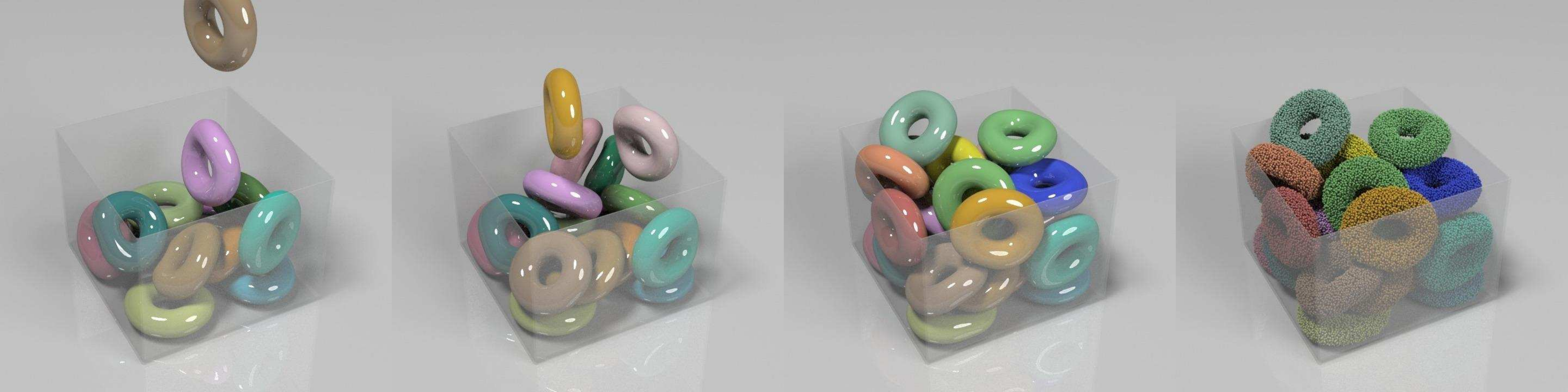}
\caption{Dropping torus with framerate $24 Hz$.\label{fig:torus-render}}
\end{center}
\end{figure}

We drop $25$ tori (with $8592$ particles each) into a box with width $0.4 \times
0.4$ and height $0.3$. Each torus has inner radius $0.03$ and outer radius $0.06$ and is
sampled at height $1.0$ with random initial rotation around the ground normal. The
material density is $5$. Young's modulus is $150$ and Poisson's ratio is $0.3$.  Figure
\ref{fig:torus-render} shows the particles and the reconstructed surfaces at time $8.50$.

\section*{Acknowledgements}
We thank Daniel Ram and Theodore Gast for their insightful suggestions.
The authors were partially supported by NSF CCF-1422795,
ONR (N000141110719, N000141210834), DOD (W81XWH-15-1-0147), Intel STC-Visual Computing
Grant (20112360) as well as a gift from Disney Research.


\bibliography{paper}

\end{document}